\documentclass[journal]{IEEEtran}
\IEEEoverridecommandlockouts
\usepackage{lineno}
\usepackage{hyperref}
\usepackage{cite}
\usepackage{amsmath,amssymb,amsfonts,cases} 
\usepackage{amsmath}
\usepackage{amsthm}
\DeclareMathOperator*{\argmax}{argmax}

\usepackage{graphicx}
\usepackage{textcomp}
\usepackage{xcolor}
\usepackage{graphicx}
\usepackage{float}
\usepackage{subfigure}
\usepackage{amsmath,mleftright}
\usepackage{amsfonts,amssymb}
\usepackage{mathrsfs}
\usepackage{mathtools}
\usepackage{algorithm}
\usepackage{algorithmic}
\usepackage{bm}
\usepackage{multirow}
\usepackage{array}
\usepackage{amssymb}
\usepackage{amsmath}
\usepackage{cite}
\usepackage{url}
\usepackage{xcolor}
\usepackage{cite,graphicx,amsmath,amssymb}
\usepackage{subfigure}
\usepackage{fancyhdr}
\usepackage{mdwmath}
\usepackage{mdwtab}
\usepackage{caption}
\usepackage{amsthm}
\usepackage{setspace}
\usepackage{bm}
\usepackage{mathtools}
\usepackage{dsfont}
\usepackage{bbm}
\usepackage{framed}
\newtheorem{remark}{Remark}
\newtheorem{theorem}{Theorem}

\newtheorem{lemma}{Lemma}

\newtheorem{corollary}{Corollary}

\makeatletter
\newcommand{\biggg}{\bBigg@{3}}
\newcommand{\Biggg}{\bBigg@{3.5}}
\makeatother
\makeatletter
\renewcommand{\maketag@@@}[1]{\hbox{\m@th\normalsize\normalfont#1}}%
\makeatother
\def\BibTeX{{\rm B\kern-.05em{\sc i\kern-.025em b}\kern-.08em
    T\kern-.1667em\lower.7ex\hbox{E}\kern-.125emX}}
    \expandafter\def\expandafter\normalsize\expandafter{%
    \normalsize%
    \setlength\abovedisplayskip{4pt}%
    \setlength\belowdisplayskip{4pt}%
    \setlength\abovedisplayshortskip{2pt}%
    \setlength\belowdisplayshortskip{2pt}%
}
\begin{document}
\title{Capacity Characterization of Pinching-Antenna Systems}
\author{Chongjun~Ouyang, Zhaolin~Wang, Yuanwei~Liu, and Zhiguo~Ding\vspace{-10pt}
\thanks{C. Ouyang and Z. Wang are with the School of Electronic Engineering and Computer Science, Queen Mary University of London, London, E1 4NS, U.K. (e-mail: \{c.ouyang, zhaolin.wang\}@qmul.ac.uk).}
\thanks{Y. Liu is with the Department of Electrical and Electronic Engineering, The University of Hong Kong, Hong Kong (email: yuanwei@hku.hk).}
\thanks{Z. Ding is with the School of Electrical and Electronic Engineering, The University of Manchester, Manchester, M13 9PL, U.K., and also with the Department of Electrical Engineering and Computer Science, Khalifa University, Abu Dhabi, UAE (e-mail: zhiguo.ding@manchester.ac.uk).}}
\maketitle
\begin{abstract}
Unlike conventional systems using a fixed-location antenna, the channel capacity of the pinching-antenna system (PASS) is determined by the activated positions of pinching antennas. This article characterizes the capacity region of multiuser PASS, where a single pinched waveguide is deployed to enable both uplink and downlink communications. The capacity region of the uplink channel is first characterized. \romannumeral1) For the single-pinch case, closed-form expressions are derived for the optimal antenna activation position, along with the corresponding capacity region and the achievable data rate regions under time-division multiple access (TDMA) and frequency-division multiple access (FDMA). It is proven that the capacity region of PASS encompasses that of conventional fixed-antenna systems, and that the FDMA rate region contains the TDMA rate region. \romannumeral2) For the multiple-pinch case, inner and outer bounds on the capacity region are derived using an element-wise alternating antenna position optimization technique and the Cauchy-Schwarz inequality, respectively. The achievable FDMA rate region is also derived using the same optimization framework, while the TDMA rate region is obtained through an antenna position refinement approach. The analysis is then extended to the downlink PASS using the uplink-downlink duality framework. It is proven that the relationships among the downlink capacity and rate regions are consistent with those in the uplink case. Numerical results demonstrate that: \romannumeral1) the derived bounds closely approximate the exact capacity region, \romannumeral2) PASS yields a significantly enlarged capacity region compared to conventional fixed-antenna systems, and \romannumeral3) in the multiple-pinch case, TDMA and FDMA are capable of approaching the channel capacity limit.
\end{abstract}
\section{Introduction}
Over the past four decades, multiple-input multiple-output (MIMO) technology has revolutionized wireless communications \cite{heath2018foundations}. More recently, research has increasingly focused on \emph{reconfigurable antennas}, which serve as an outer layer complementing conventional digital and/or analog beamforming \cite{heath2025tri}. These antennas enable \emph{electromagnetic (EM) beamforming} by dynamically adjusting EM properties such as polarization, operating frequency, and radiation patterns \cite{heath2025tri}. This capability allows for flexible channel manipulation with minimal energy consumption so as to enhance system performance without significantly increasing radio-frequency (RF) complexity or power usage \cite{heath2025tri,castellanos2023energy}. Reconfigurable antennas are now considered as a critical component of the emerging \emph{tri-hybrid beamforming} architecture that is poised to reshape the sixth-generation (6G) wireless networks \cite{heath2025tri,castellanos2023energy,castellanos2025embracing}.

Typical reconfigurable antennas include dynamic metasurface antennas (DMAs) \cite{shlezinger2021dynamic}, reconfigurable intelligent surfaces (RISs) \cite{liu2021reconfigurable}, fluid antennas \cite{wong2020fluid}, and movable antennas \cite{zhu2024movable}. DMAs and RISs leverage recent advances in metamaterials to manipulate EM radiation or reflection via programmable meta-atoms. These can dynamically adjust parameters such as resonant frequency, polarization, radiation patterns, and beam direction. In contrast, fluid and movable antennas aim to optimize antenna positions to exploit spatial diversity and mitigate small-scale fading. Despite their promising performance improvements \cite{zhang2021intelligent,castellanos2023energy,castellanos2025embracing,new2024tutorial,zhu2025tutorial}, these technologies face key limitations, particularly regarding \emph{free-space path loss} and \emph{line-of-sight (LoS) blockage}. Their ability to reconfigure the wireless channel is typically constrained to apertures spanning only a few to perhaps tens of wavelengths. This limited spatial reconfiguration is insufficient to mitigate large-scale path loss or blind spots for distant or cell-edge users. Furthermore, in these systems, it is costly to change their antenna configuration, i.e., add or remove antennas, once these systems are established.
\subsection{Pinching Antennas}
To address the limitations of conventional reconfigurable-antenna technologies, NTT DOCOMO has introduced a novel architecture known as the \emph{pinching antenna} \cite{pinching_antenna1,suzuki2022pinching}. This design utilizes a dielectric waveguide \cite{pozar2021microwave} as the transmission medium, with EM waves radiated via small dielectric particles affixed at specific positions along the waveguide \cite{suzuki2022pinching}. These radiating elements---referred to as pinching antennas---are typically mounted on the ends of plastic clips, which allows them to be attached or detached in a manner reminiscent of clothespins on a clothesline \cite{yang2025pinching,liu2025pinching}. Each pinching antenna can be independently activated or deactivated, which enables real-time reconfiguration of the array geometry as well as the wireless channel. This mechanism facilitates a highly flexible and scalable deployment strategy, which we refer to as \emph{pinching beamforming} \cite{liu2025pinching}. 

Unlike conventional reconfigurable-antenna systems, the waveguide in the \emph{pinching-antenna system (PASS)} can be extended to arbitrary lengths, which allows the deployment of antennas in close proximity to users and thereby establishing robust LoS connections. This spatial adaptability enhances link quality by mitigating large-scale path loss and eliminating coverage holes and LoS blockages \cite{suzuki2022pinching}. PASS is also highly cost-effective and easy to deploy, as it requires only the addition or removal of dielectric materials, without the need for complex radio-frequency circuitry. 

In essence, PASS can be regarded as a practical realization of fluid- or movable-antenna systems \cite{new2024tutorial,zhu2025tutorial}, but with enhanced flexibility and scalability compared to traditional architectures. In recognition of NTT DOCOMO's original contribution \cite{pinching_antenna1,suzuki2022pinching}, we adopt the term \emph{``PASS''} throughout this article. Furthermore, PASS is well aligned with the emerging vision of surface-wave communication superhighways \cite{wong2020vision}, which aim to exploit in-waveguide propagation through reconfigurable waveguides to minimize path loss and enhance signal delivery efficiency \cite{liu2024path,chu2024propagation}.

Owing to its distinctive characteristics, PASS-assisted communications have attracted increasing research interest. Initial theoretical studies analyzed the ergodic rate of PASS to quantify its performance gains over conventional fixed-antenna systems \cite{ding2024flexible}. Building upon this foundation, subsequent work studied the achievable array gain and identified the optimal number of pinching antennas required to maximize performance \cite{ouyang2025array}. Incorporating LoS blockage into the analysis, further studies evaluated both the ergodic rate and outage probability achieved by PASS \cite{tyrovolas2025performance,ding2025blockage}. These results demonstrated that PASS effectively mitigates large-scale path loss and LoS blockage, outperforming not only traditional fixed-antenna systems but also existing fluid and movable antenna designs. In parallel, several pinching beamforming algorithms have been developed to optimize the spatial placement of active pinching antennas along the waveguide, with the aim of maximizing overall communication performance \cite{xu2024rate,wang2024antenna,tegos2024minimum,wang2025modeling,bereyhi2025downlink,bereyhi2025mimo,xie2025low}. In addition to communications, initial efforts have also explored the potential of PASS to wireless sensing, including user localization \cite{ding2025pinching,bozanis2025CRLB,wang2025wireless}, target detection \cite{zhang2025integrated}, and joint communications and sensing \cite{ouyang2025rate}.
\subsection{Motivation and Contributions}
Despite recent progress, a fundamental question of PASS remains unanswered: \emph{its information-theoretic channel capacity limits are not yet fully understood}. Addressing this gap is critical to quantifying the maximum achievable performance gains enabled by PASS. While initial efforts have discussed the single-user case \cite{ding2024flexible,ouyang2025array,xu2024rate}, the multiuser setting, where performance is characterized by the \emph{capacity region (the set of all achievable rate-tuples)}, remains largely unexplored. It is important to note that the capacity regions of conventional multiuser channels have been extensively investigated. What fundamentally distinguishes PASS is that its capacity region depends on the activated positions of pinching antennas, making the capacity analysis inherently more complex. Furthermore, the achievable data rate regions under practical orthogonal multiple access (OMA) schemes, including time-division multiple access (TDMA) and frequency-division multiple access (FDMA), have not been analyzed for PASS. Last, but not least, the relationship between these achievable regions and the capacity region is still unclear. 

To bridge these gaps and examine how pinching beamforming influences capacity limits, this article investigates a two-user communication system enabled by a single pinched waveguide. Our aim is to derive the capacity region and characterize the TDMA and FDMA achievable data rate regions for both the \emph{uplink multiple-access channel (MAC)} and \emph{downlink broadcast channel (BC)}. The extension to systems with more than two users is left for future work. The main contributions of this article are summarized as follows.
\begin{itemize}
  \item First, we investigate the PASS-enabled uplink channel with a single pinching antenna. To characterize the single-pinch uplink capacity region, we employ the \emph{rate-profile approach} \cite{zhang2010cooperative} by solving a series of sum-rate maximization problems through optimization of the activated pinching antenna position. We derive closed-form \emph{optimal} solutions for each user rate-profile ratio and obtain the complete capacity region. We further derive a closed-form expression for the achievable data rate region under TDMA and apply the rate-profile method to characterize the rate region with FDMA. On this basis, we prove that: \romannumeral1) the capacity region achieved by a conventional fixed-antenna system is nested in that of the PASS, and \romannumeral2) the FDMA rate region contains the TDMA rate region.
  \item Next, we extend our analysis to the more general multiple-pinch case, where multiple pinching antennas are deployed along the waveguide to serve the uplink users. Using the rate-profile approach, we propose an efficient element-wise alternating optimization algorithm to design the pinching beamforming. By further using the time-sharing strategy, we compute an \emph{achievable inner bound} of the uplink capacity region. To complement this, we derive a \emph{capacity outer bound} based on the Cauchy-Schwarz inequality. For TDMA, we employ an antenna position refinement method to optimize the pinching beamforming and derive the rate region. For FDMA, we apply the element-wise alternating optimization method to obtain an achievable rate region. Furthermore, we analyze the scaling behavior of the capacity region with respect to the number of pinching antennas and demonstrate the existence of an optimal antenna number that maximizes the capacity region.
  \item Then, we extend the results for the uplink PASS to the downlink scenario by leveraging the celebrated \emph{uplink-downlink duality} framework. Based on the capacity and rate regions derived for the dual uplink PASS, we develop computationally efficient methods to characterize the downlink capacity region as well as the achievable TDMA and FDMA rate regions in both the single-pinch and multiple-pinch cases. Furthermore, we analytically prove that the inclusion relationships among the capacity and achievable rate regions observed in the uplink setting also hold for the downlink PASS.
  \item Finally, we present numerical results to validate the theoretical analysis and assess the tightness of the proposed capacity bounds. The results demonstrate that \romannumeral1) the proposed inner and outer bounds closely approximate the true capacity region in the multiple-pinch scenario, \romannumeral2) PASS significantly enlarges the capacity region as well as the TDMA and FDMA achievable rate regions compared to conventional fixed-antenna systems in both uplink and downlink channels, and \romannumeral3) for the multiple-pinch case, practical OMA schemes---namely TDMA and FDMA---can \emph{nearly achieve} the full capacity region in both uplink and downlink channels. These findings underscore the scalability and effectiveness of PASS in enlarging the capacity region of wireless multiuser systems.
\end{itemize}

The remainder of this paper is organized as follows. Section {\ref{Section: System Model}} presents the system model for the PASS-enabled multiuser channels. Sections \ref{Section: PASS-Enabled Uplink Channels} and \ref{Section: PASS-Enabled Downlink Channels} characterize the capacity region and the TDMA/FDMA rate regions for the uplink and downlink channels, respectively. Section \ref{Section_Numerical_Results} provides numerical results along with detailed performance evaluations and discussions. Finally, Section \ref{Section_Conclusion} concludes the paper.
\subsubsection*{Notations}
Throughout this paper, scalars and vectors are denoted by non-bold and bold lower-case letters, respectively. The notations $\lvert a\rvert$ and $\lVert \mathbf{a} \rVert$ represent the magnitude of scalar $a$ and the norm of vector $\mathbf{a}$, respectively. The transpose operator is denoted by $(\cdot)^{\mathsf{T}}$. The statistical expectation operator is denoted by ${\mathbbmss{E}}\{\cdot\}$. The notation ${\mathcal{CN}}({\bm\mu},\mathbf{X})$ refers to the circularly symmetric complex Gaussian distribution with mean $\bm\mu$ and covariance matrix $\mathbf{X}$. The sets ${\mathbbmss{C}}$ and ${\mathbbmss{R}}$ denote the complex and real spaces, respectively. The convex hull of a set is denoted by $\rm{Conv}(\cdot)$, and the union operation is denoted by $\bigcup$. The big-O notation is given by ${\mathcal{O}}(\cdot)$. 
\section{System Model}\label{Section: System Model}
\begin{figure}[!t]
\centering
\includegraphics[width=0.4\textwidth]{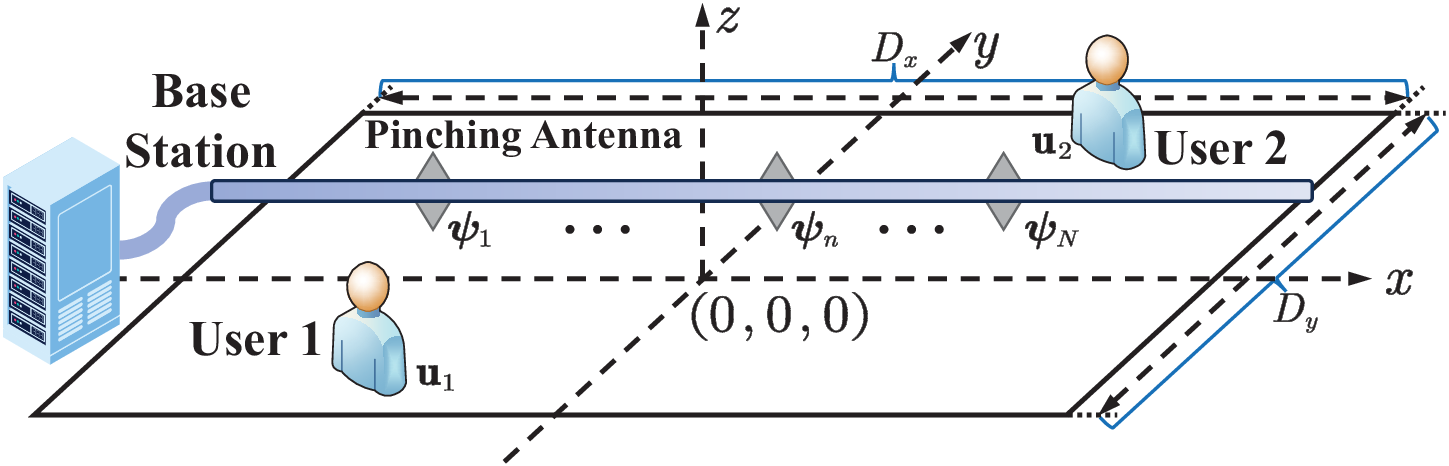}
\caption{Illustration of a multiuser PASS.}
\label{Figure1}
\vspace{-15pt}
\end{figure}

Consider a communication network where one base station (BS) employs a single pinched waveguide to serve two single-antenna users, as illustrated in {\figurename} {\ref{Figure1}}. In the uplink, each user transmits an independent message to the BS; while in the downlink, the BS delivers independent messages to the users. A total of $N$ pinching antennas are activated on the waveguide to enhance user communication rates through \emph{pinching beamforming}. %Note that the PASS-enabled uplink (from the users to the BS) and downlink (from the BS to the users) transmissions are similar to the conventional discrete memoryless uplink and downlink channels \cite{el2011network}, but with the user effective channels controllable by the PASS.

As illustrated in {\figurename} {\ref{Figure1}}, the waveguide is aligned along the $x$-axis at a height $d$. To analyze the theoretical performance limits of PASS-enabled multiuser communications, we adopt a free-space LoS channel model. This choice is motivated by the suitability of PASS in high-frequency bands \cite{suzuki2022pinching}, where LoS propagation is generally dominant \cite{ouyang2024primer}. The effects of multipath fading \cite{xiao2025channel} are not considered in this analysis and are left for future investigation. Under the LoS assumption, the spatial channel coefficient between the $n$th pinching antenna and user $k\in\{1,2\}\triangleq{\mathcal{K}}$ is given by \cite{liu2023near}:
\begin{align}\label{Spatial_Channel_Model}
h({\mathbf{u}}_{k},{\bm\psi}_n)\triangleq\frac{\eta^{\frac{1}{2}}{\rm{e}}^{-{\rm{j}}k_0\lVert{\mathbf{u}}_{k}-{\bm\psi}_n\rVert}}{\lVert{\mathbf{u}}_{k}-{\bm\psi}_n\rVert},
\end{align}
where $\eta\triangleq\frac{c^2}{16\pi^2f_{\rm{c}}^2}$, $k_0=\frac{2\pi}{\lambda}$, $c$ denotes the speed of light, $f_{\rm{c}}$ is the carrier frequency, and $\lambda$ is the free-space wavelength. The location of user $k$ is denoted by ${\mathbf{u}}_{k}\triangleq[x_{k},y_{k},0]^{\mathsf{T}}$ for $k\in{\mathcal{K}}$, and the position of the $n$th pinching antenna is given by ${\bm\psi}_n\triangleq[q_n,y_{\rm{p}},d]^{\mathsf{T}}$ for $n\in{\mathcal{N}}\triangleq\{1,\ldots,N\}$. Without loss of generality, the antenna deployment is assumed to satisfy
\begin{align}
q_{\max}\geq q_{n}>q_{n'},~\forall n>n', 
\end{align}
where $q_{\max}$ denotes the maximum deployment range of the pinching antennas. Besides, we assume that $x_1\leq x_2$.
\subsection{Uplink Transmission with PASS}
Let $s_k\in{\mathbbmss{C}}$ denote the desired information symbol for user $k\in{\mathcal{K}}$ with zero mean and unit variance. The symbols $s_k$'s are mutually independent across users. The transmitted signal by user $k$ is modeled as $x_k=\sqrt{p_k}s_k$, which satisfies ${\mathbbmss{E}}\{\lvert x_k\rvert^2\}=p_k\leq P_k$, with $p_k$ denoting the transmit power of user $k$ and $P_k$ denoting its maximum value. The received signal at the BS is given by
\begin{subequations}\label{Signal_Model_Uplink}
\begin{align}
{{y}}&=\sum\nolimits_{n=1}^{N}{\rm{e}}^{-{\rm{j}}{\phi_n}}\left(\sum\nolimits_{k=1}^{2}h({\mathbf{u}}_{k},{\bm\psi}_n)x_k+z_n^{\rm{U}}\right)\\
&=\sum\nolimits_{k=1}^{2}\sqrt{p_k}{\mathbf{h}}^{\mathsf{T}}({\mathbf{u}}_{k},{\mathbf{q}}){\bm\phi}{{s}}_k+z_{\rm{U}},
\end{align}
\end{subequations}
where $z_n^{\rm{U}}\sim{\mathcal{CN}}(0,\sigma^2)$ represents the additive Gaussian noise at the $n$th pinching antenna with $\sigma^2$ being the noise power,
\begin{subequations}
\begin{align}
{\bm\phi}&\triangleq[{\rm{e}}^{-{\rm{j}}\phi_{1}},\ldots,{\rm{e}}^{-{\rm{j}}{\phi_{N}}}]^{\mathsf{T}}\in{\mathbbmss{C}}^{N\times1},\\
{\mathbf{h}}({\mathbf{u}}_{k},{\mathbf{q}})&\triangleq[h({\mathbf{u}}_{k},{\bm\psi}_1),\ldots,h({\mathbf{u}}_{k},{\bm\psi}_N)]^{\mathsf{T}}\in{\mathbbmss{C}}^{N\times1},\\
{\mathbf{q}}&\triangleq[q_1,\ldots,q_N]^{\mathsf{T}}\in{\mathbbmss{R}}^{N\times1},
\end{align}
\end{subequations}
and $z_{\rm{U}}\triangleq\sum_{n=1}^{N}{\rm{e}}^{-{\rm{j}}{\phi_n}}z_n^{\rm{U}}\sim{\mathcal{CN}}(0,N\sigma_{\rm{U}}^2)$. The term 
\begin{align}
\phi_n\triangleq\frac{2\pi\lVert{\bm\psi}_n-{\bm\psi}_0\rVert}{\lambda_{\rm{g}}}=\frac{2\pi\lvert q_n-q_0\rvert}{\lambda_{\rm{g}}},\quad n\in\mathcal{N}, 
\end{align}
denotes the in-waveguide phase shift for the $n$th pinching antenna, where ${\bm\psi}_0\triangleq[q_0,y_{\rm{p}},d]^{\mathsf{T}}$ is the location of the waveguide feed point with $q_0\leq q_{1}$ and $[x_1,x_2]\subseteq[q_0,q_{\max}]$, $\lambda_{\rm{g}}=\frac{\lambda}{n_{\rm{eff}}}$ is the guided wavelength, and $n_{\rm{eff}}$ is the effective refractive index of the dielectric waveguide \cite{pozar2021microwave}. For each user $k\in{\mathcal{K}}$, let $R_k^{\rm{U}}$ denote its achievable uplink rate.
\subsubsection*{Conventional Fixed-Antenna System}
For comparison, we also consider the uplink signal model in a conventional fixed-antenna system, where a single antenna\footnote{We assume that the conventional-antenna system employs only a single transmit antenna, not only due to the high cost of deploying additional antennas, but also because such systems inherently lack such flexibility afforded by the PASS architecture.} is placed at ${\bm\psi}_{\rm{f}}\triangleq[q_{\rm{f}},y_{\rm{p}},d]^{\mathsf{T}}$ with $q_{\rm{f}}\in[q_0,q_{\max}]$. The received signal at the BS can be expressed as follows:
\begin{align}\label{CASS_Uplink_Model}
y_{\rm{f}}=\sum\nolimits_{k=1}^{2}\sqrt{p_k}h_k^{\rm{f}}s_k + z_{{\rm{U}}}^{\rm{f}},
\end{align}
where $z_{{\rm{U}}}^{\rm{f}}\sim{\mathcal{CN}}(0,\sigma^2)$ denotes the additive Gaussian noise, and $h_k^{\rm{f}}\triangleq h({\mathbf{u}}_{k},{\bm\psi}_{\rm{f}})$ represents the spatial channel between the BS and user $k$. By comparing \eqref{CASS_Uplink_Model} with the PASS-enabled model in \eqref{Signal_Model_Uplink}, we observe that the received signal under PASS is influenced by the spatial configuration of the pinching antennas, namely the \emph{pinching beamformer} $\mathbf{q}=[q_1,\ldots,q_N]^{\mathsf{T}}$. This additional spatial degree of freedom allows PASS to dynamically adapt its effective channel, thereby enabling more stable and stronger LoS links to users. In contrast, the conventional fixed-antenna system lacks this reconfigurability and remains limited in its ability to optimize the wireless channel.
\subsection{Downlink Transmission with PASS}
For ease of exposition, we consider a dual-channel setup in which all downlink channels are assumed to be identical to their uplink counterparts. Under this assumption, the received signal at user $k$ can be written as follows:
\begin{align}\label{Signal_Model_Downlink}
y_k=\sqrt{1/N}{\mathbf{h}}^{\mathsf{T}}({\mathbf{u}}_{k},{\mathbf{q}}){\bm\phi}{{x}}+z_{k},\quad k\in{\mathcal{K}},
\end{align}
where $x=\sqrt{p_1}s_1+\sqrt{p_2}s_2$ denotes the transmitted signal from the BS, with $p_k$ and $s_k$ representing the transmit power and information symbol for user $k$, respectively; $z_k\sim{\mathcal{CN}}(0,\sigma^2)$ denotes the receiver noise at user $k$. We consider a transmit power budget $P$ at the BS, which yields $p_1+p_2\leq P$. It is further assumed that the total power is equally distributed among the $N$ active pinching antennas \cite{ding2024flexible,wang2025modeling}, which results in the $\sqrt{1/N}$ scaling factor in \eqref{Signal_Model_Downlink}. For each user $k\in{\mathcal{K}}$, let $R_k^{\rm{D}}$ denote its achievable downlink rate.
\subsubsection*{Conventional Fixed-Antenna System}
For comparison, we also consider a conventional fixed-antenna downlink channel, where a single antenna is located at ${\bm\psi}_{\rm{f}}=[q_{\rm{f}},y_{\rm{p}},d]^{\mathsf{T}}$. The received signal at user $k$ is given by
\begin{align}\label{CASS_Downlink_Model}
y_k^{\rm{f}}=h_k^{\rm{f}}x + z_k.
\end{align}
%Comparing \eqref{CASS_Downlink_Model} with the downlink model in \eqref{Signal_Model_Downlink}, we note that the performance in PASS depends on the design of pinching beamformer $\mathbf{q}$. This suggests that PASS can potentially yields a broader capacity region than conventional-antenna systems.

In the following, we characterize the capacity region of the PASS-enabled two-user uplink channel, which constitutes all the achievable rate tuples $(R_1^{\rm{U}},R_2^{\rm{U}})$. We also derive the achievable data rate regions under practical OMA schemes, including TDMA and FDMA. The results are subsequently extended to the PASS-enabled downlink channel by applying the celebrated \emph{uplink-downlink duality} framework \cite{el2011network}.
\section{PASS-Enabled Uplink Channels}\label{Section: PASS-Enabled Uplink Channels}
This section characterizes the capacity region and the achievable data rate regions under TDMA and FDMA for the PASS-enabled two-user uplink channel. 
\subsection{Overview of the Capacity Region}
\begin{figure}[!t]
\centering
\includegraphics[width=0.3\textwidth]{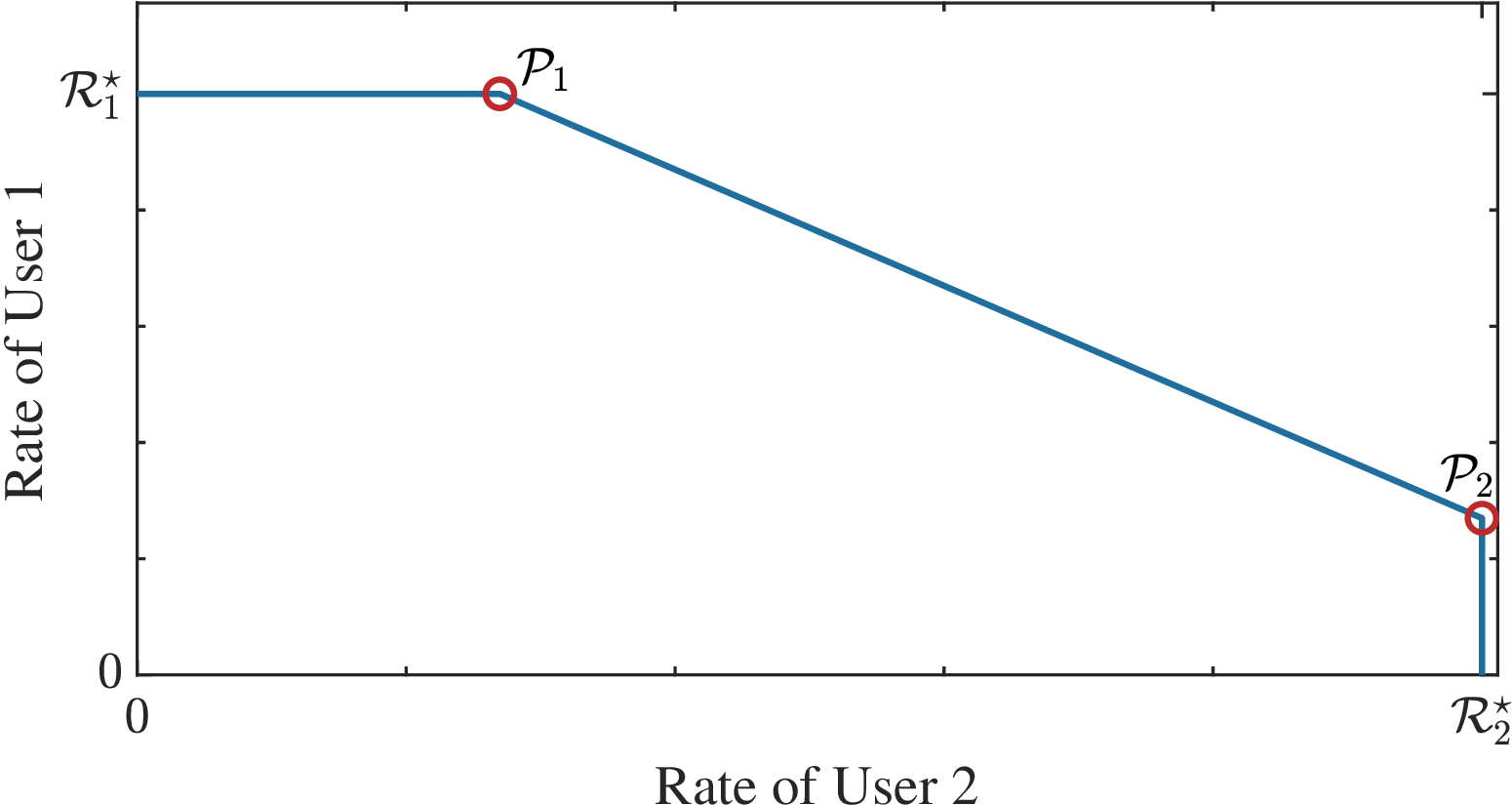}
\caption{An illustrated example of the capacity region for a two-user uplink channel.}
\label{Figure_MAC_Capacity_Region_AWGN}
\vspace{-15pt}
\end{figure}

As a benchmark, we first present the capacity region of the conventional fixed-antenna system, as described by \eqref{CASS_Uplink_Model}. This region is known to form a \emph{pentagon}, as illustrated in {\figurename} {\ref{Figure_MAC_Capacity_Region_AWGN}}, and consists of all rate pairs $(R_1^{\rm{U}},R_2^{\rm{U}})$ that satisfy the following constraints \cite{el2011network}:
\begin{subequations}\label{CASS_Uplink_Capacity_Region}
\begin{align}
R_k^{\rm{U}}&\leq \log_2(1+P_k\lvert h_k^{\rm{f}}\rvert^2/\sigma^2),\quad k\in{\mathcal{K}},\label{CASS_Uplink_Capacity_Region_Constraint1}\\
R_1^{\rm{U}}+R_2^{\rm{U}}&\leq \log_2(1+P_1\lvert h_1^{\rm{f}}\rvert^2/\sigma^2+P_2\lvert h_2^{\rm{f}}\rvert^2/\sigma^2).
\label{CASS_Uplink_Capacity_Region_Constraint2}
\end{align}
\end{subequations}
The corresponding capacity region is defined as follows:
\begin{align}\label{Conventional_Fixed_Antenna_Uplink}
{\mathcal{C}}_{\rm{f}}^{\rm{U}}\triangleq\{(R_1^{\rm{U}},R_2^{\rm{U}})\left\lvert R_1^{\rm{U}}\geq0,R_2^{\rm{U}}\geq0,\eqref{CASS_Uplink_Capacity_Region_Constraint1},\eqref{CASS_Uplink_Capacity_Region_Constraint2}\right.\}.
\end{align}

Achieving the capacity region ${\mathcal{C}}_{\rm{f}}^{\rm{U}}$ requires the use of \emph{successive interference cancellation (SIC)} or \emph{joint decoding} at the BS \cite{el2011network}. Under SIC, the BS first decodes one user's message while treating the other user's signal as interference. After decoding and removing the first user's signal, the BS proceeds to decode the remaining message. Let $\bm\pi\triangleq[\pi_1,\pi_2]^{\mathsf{T}}$ denote the decoding order, with ${\bm\pi}=[2,1]^{\mathsf{T}}\triangleq{\bm\pi}^{\rm{\uppercase\expandafter{\romannumeral1}}}$ indicating that user $1$ is decoded after user $2$, and ${\bm\pi}=[1,2]^{\mathsf{T}}\triangleq{\bm\pi}^{\rm{\uppercase\expandafter{\romannumeral2}}}$ otherwise. As shown in {\figurename} {\ref{Figure_MAC_Capacity_Region_AWGN}}, the corner points ${\mathcal{P}}_1$ and ${\mathcal{P}}_2$ are achieved using SIC with decoding orders ${\bm\pi}^{\rm{\uppercase\expandafter{\romannumeral1}}}$ and ${\bm\pi}^{\rm{\uppercase\expandafter{\romannumeral2}}}$, respectively, whereas the line segment connecting ${\mathcal{P}}_1$ and ${\mathcal{P}}_2$ is achieved via \emph{time sharing} between ${\mathcal{P}}_1$ and ${\mathcal{P}}_2$. It is also worth noting that ${\mathcal{C}}_{\rm{f}}^{\rm{U}}$ is achieved when each user transmits at its maximum power, i.e., $p_k=P_k$ for $k\in{\mathcal{K}}$ \cite{el2011network}.

In the uplink PASS, for a given pinching beamformer $\mathbf{q}$ and transmit power vector ${\mathbf{p}}\triangleq[p_{\pi_1},p_{\pi_2}]^{\mathsf{T}}$, the achievable rates of the first and second decoded users under a SIC decoding order $\bm\pi$ are given by
\begin{subequations}
\begin{align}
{\mathcal{R}}_{1}^{\rm{U}}({\bm\pi},{\mathbf{q}},{\mathbf{p}})&\triangleq\log_2\left(1+\frac{p_{\pi_1}\lvert h_{\pi_1}({\mathbf{q}})\rvert^2}{p_{\pi_2}\lvert h_{\pi_2}({\mathbf{q}})\rvert^2+\sigma^2}\right),\\
{\mathcal{R}}_{2}^{\rm{U}}({\bm\pi},{\mathbf{q}},{\mathbf{p}})&\triangleq\log_2\left(1+\frac{p_{\pi_2}\lvert h_{\pi_2}({\mathbf{q}})\rvert^2}{\sigma^2}\right),
\end{align}
\end{subequations}
respectively, where $h_k({\mathbf{q}})\triangleq\frac{1}{\sqrt{N}}{\mathbf{h}}^{\mathsf{T}}({\mathbf{u}}_{k},{\mathbf{q}}){\bm\phi}$ denotes the effective channel gain for user $k\in\{1,2\}$. Following the derivations in \eqref{CASS_Uplink_Capacity_Region}, for a given $\mathbf{q}$, all achievable rate pairs $(R_1^{\rm{U}},R_2^{\rm{U}})$ under the power budget constraints $p_k\leq P_k$ ($k\in{\mathcal{K}}$) satisfy the following:
\begin{subequations}\label{PASS_Uplink_Capacity_Region_Constraint}
\begin{align}
&R_k^{\rm{U}}\leq \log_2(1+P_k\lvert h_k({\mathbf{q}})\rvert^2/\sigma^2)\triangleq r_k^{\rm{U}}({\mathbf{q}}),k\in{\mathcal{K}}, \label{PASS_Uplink_Capacity_Region_Constraint1}\\
&R_1^{\rm{U}}+R_2^{\rm{U}}\leq \log_2(1+(P_1\lvert h_1({\mathbf{q}})\rvert^2
+P_2\lvert h_2({\mathbf{q}})\rvert^2)/\sigma^2)\nonumber\\
&\qquad\qquad\triangleq r_{1,2}^{\rm{U}}({\mathbf{q}}).
\label{PASS_Uplink_Capacity_Region_Constraint2}
\end{align}
\end{subequations}
The capacity region is given by
\begin{align}\label{PASS_Uplink_Capacity_Region_Specific}
{\mathcal{C}}^{\rm{U}}({\mathbf{q}})\triangleq\{(R_1^{\rm{U}},R_2^{\rm{U}})\left\lvert R_1^{\rm{U}}\geq0,R_2^{\rm{U}}\geq0,\eqref{PASS_Uplink_Capacity_Region_Constraint1},\eqref{PASS_Uplink_Capacity_Region_Constraint2}\right.\}.
\end{align}

By flexibly designing the pinching beamformer ${\mathbf{q}}$, the system can achieve any rate pair within the union of the sets ${\mathcal{C}}^{\rm{U}}({\mathbf{q}})$ across all feasible ${\mathbf{q}}$'s. Moreover, by incorporating \emph{time sharing} among different $\mathbf{q}$'s, the capacity region of the PASS-enabled uplink channel is given by the convex hull of such a union set \cite{goldsmith2003capacity}:
\begin{align}\label{PASS_Uplink_Capacity_Region_General_Exhausitive}
{\mathcal{C}}^{\rm{U}}\triangleq{\rm{Conv}}\Big(\bigcup\nolimits_{{\mathbf{q}}\in{\mathcal{P}}}{\mathcal{C}}^{\rm{U}}({\mathbf{q}})\Big),
\end{align}
where ${\mathcal{P}}\triangleq\{\left.[q_1,\ldots,q_N]^{\mathsf{T}}\in{\mathbbmss{R}}^{N\times1}\right\rvert \lvert q_n-q_{n'}\rvert\geq\Delta,n\ne n'\}$ denotes the feasible set of ${\mathbf{q}}$, with $\Delta$ being the minimum inter-antenna spacing to mitigate EM mutual coupling \cite{ivrlavc2010toward}. 

Although ${\mathcal{C}}^{\rm{U}}$ can be computed by evaluating all ${\mathcal{C}}^{\rm{U}}({\mathbf{q}})$'s over ${\mathcal{P}}$ and taking their convex hull, this procedure is computationally intensive. To reduce the complexity, we introduce an alternative method to characterize ${\mathcal{C}}^{\rm{U}}$ more efficiently.
\subsection{Rate-Profile Based Capacity Region Characterization}
Recall that for each beamformer $\mathbf{q}$, all achievable rate pairs on the \emph{frontier}, or \emph{Pareto boundary}, of the capacity region ${\mathcal{C}}^{\rm{U}}({\mathbf{q}})$---except those requiring \emph{time sharing} between the two corner points---can be attained through SIC at the BS. Motivated by this, we propose to first characterize the union of SIC-achievable Pareto-optimal rate pairs across all feasible ${\mathbf{q}}\in{\mathcal{P}}$, and then apply \emph{time sharing} among these Pareto-optimal points to construct the complete uplink capacity region. 

To address the first task, we adopt the \emph{rate-profile} approach from \cite{zhang2010cooperative,zhang2021intelligent}. This method serves as an effective tool for characterizing the capacity region by formulating and solving a sum-rate maximization problem, which yields all rate pairs along the frontier of the capacity region; an illustrative explanation is provided in \cite[{\figurename} 1]{zhang2010cooperative}. Specifically, for a given decoding order $\bm\pi$, each rate pair located on the Pareto boundary of the capacity region can be determined by solving the following sum-rate maximization problem \cite{zhang2010cooperative,zhang2021intelligent}:
\begin{subequations}\label{Rate_Profile_Uplink_General}
\begin{align}
\max_{{\mathbf{q}},R,p_1,p_2}~&R\\
{\rm{s.t.}}~&{\mathbf{q}}\in{\mathcal{P}},R\geq0,p_{\pi_k}\in[0,P_{\pi_k}],k\in{\mathcal{K}},\\
&{\mathcal{R}}_{1}^{\rm{U}}({\bm\pi},{\mathbf{q}},{\mathbf{p}})\geq \alpha R,{\mathcal{R}}_{2}^{\rm{U}}({\bm\pi},{\mathbf{q}},{\mathbf{p}})\geq \overline{\alpha} R,
\label{Rate_Profile_Uplink_General_Constraint2}
\end{align}
\end{subequations}
where $\alpha\in[0,1]$ is the rate-profile factor, which specifies the rate ratio between the first decoded user and the users' sum-rate $R$, and $\overline{\alpha}\triangleq1-\alpha\in[0,1]$. The above sum-rate maximization can equivalently be reformulated as a max-min optimization of $\min\{\alpha^{-1}{\mathcal{R}}_{1}^{\rm{U}}({\bm\pi},{\mathbf{q}},{\mathbf{p}}),{\overline{\alpha}}^{-1}{\mathcal{R}}_{2}^{\rm{U}}({\bm\pi},{\mathbf{q}},{\mathbf{p}})\}$, which plays a critical role in solving the rate-profile problems formulated in this article. Referring to \eqref{PASS_Uplink_Capacity_Region_Constraint}, \eqref{PASS_Uplink_Capacity_Region_Specific}, and \eqref{PASS_Uplink_Capacity_Region_General_Exhausitive}, we establish the following result.
\vspace{-5pt}
\begin{lemma}\label{Lemma_Power_Condition}
The rate pairs on the frontier of the uplink capacity region ${\mathcal{C}}^{\rm{U}}$ are achieved when each user transmits at its maximum power, i.e., $p_{k}=P_k$ for $k\in{\mathcal{K}}$.
\end{lemma}
\vspace{-5pt}
\begin{IEEEproof}
This lemma can be readily proven by leveraging the monotonicity of $r_k^{\rm{U}}({\mathbf{q}})$ ($\forall k$) and $r_{1,2}^{\rm{U}}({\mathbf{q}})$ with respect to the transmit powers $(P_1,P_2)$, as per \eqref{PASS_Uplink_Capacity_Region_Constraint1} and \eqref{PASS_Uplink_Capacity_Region_Constraint2}.
\end{IEEEproof}
By Lemma \ref{Lemma_Power_Condition}, problem \eqref{Rate_Profile_Uplink_General} simplifies to the following:
\begin{align}\label{Rate_Profile_Uplink_Simplified}
\max_{{\mathbf{q}},R}~R\quad{\rm{s.t.}}~{\mathbf{q}}\in{\mathcal{P}},R\geq0,\eqref{Rate_Profile_Uplink_General_Constraint2},p_{k}=P_k,k\in{\mathcal{K}}.
\end{align}
Let ${\mathbf{q}}_{{\bm\pi}}^{\alpha}$ denote the optimal solution of ${\mathbf{q}}$ to problem \eqref{Rate_Profile_Uplink_Simplified} for a given $\{{\bm\pi},{\alpha}\}$, and let ${\mathcal{R}}_{\pi_k}^{\rm{U}}({\mathbf{q}}_{{\bm\pi}}^{\alpha})$ denote the corresponding achievable rate of the $k$th decoded user. Therefore, the capacity region of the uplink PASS can be characterized as follows:
\begin{align}\label{Uplink_Channel_Capacity_Region_Rate_Profile_Basic}
{\mathcal{C}}^{\rm{U}}={\rm{Conv}}\Big(\bigcup\nolimits_{\alpha\in[0,1],{\bm\pi}\in\{{\bm\pi}^{\rm{\uppercase\expandafter{\romannumeral1}}},
{\bm\pi}^{\rm{\uppercase\expandafter{\romannumeral2}}}\}}{\mathcal{C}}_{{\bm\pi}}^{\rm{U}}({\mathbf{q}}_{{\bm\pi}}^{\alpha})\Big),
\end{align}
where ${\mathcal{C}}_{{\bm\pi}}^{\rm{U}}({\mathbf{q}}_{{\bm\pi}}^{\alpha})\triangleq\{(R_1^{\rm{U}},R_2^{\rm{U}})\left\lvert R_k^{\rm{U}}\in[0,{\mathcal{R}}_{k}^{\rm{U}}({\mathbf{q}}_{{\bm\pi}}^{\alpha})],k\in{\mathcal{K}}\right.\}$.

Equation \eqref{Uplink_Channel_Capacity_Region_Rate_Profile_Basic} offers an alternative and tractable characterization of the uplink capacity region ${\mathcal{C}}^{\rm{U}}$ via \emph{rate-profile optimization}. However, problem \eqref{Rate_Profile_Uplink_Simplified} is non-convex and NP-hard. Solving it optimally is therefore challenging. To gain key design insights and reduce complexity, we begin by analyzing the single-pinch case with $N=1$, and later extend the analysis to the general multiple-pinch case with $N>1$. 
\subsection{Single-Pinch Case}
\subsubsection{Capacity Region}\label{Section: Uplink: Single-Pinch Case: Capacity Region}
When $N=1$, the effective channel gain for user $k\in{\mathcal{K}}$ reduces to:
\begin{align}\label{Uplink_SU_Channel_Gain_Basic}
\lvert h_k(q_1)\rvert=\eta^{\frac{1}{2}}\frac{1}{\sqrt{d_k^2+(x_k-q_1)^2}},
\end{align}
where $d_k^2\triangleq d^2 + (y_k-y_{\rm{p}})^2$. Inserting this into \eqref{PASS_Uplink_Capacity_Region_General_Exhausitive} gives
\begin{align}\label{Single_Pinch_Exhausitive_Search_Capacity}
{\mathcal{C}}^{\rm{U}}={\rm{Conv}}\Big(\bigcup\nolimits_{q_1\in[q_0,q_{\max}]}{\mathcal{C}}^{\rm{U}}(q_1)\Big).
\end{align}
According to \eqref{Single_Pinch_Exhausitive_Search_Capacity}, we establish the following result.
\vspace{-5pt}
\begin{lemma}\label{Lemma_Single_PA_Basic_Step}
When $N=1$, the capacity region of the two-user uplink PASS is given by 
\begin{align}\label{Single_Pinch_Single_PA_Basic_Step}
{\mathcal{C}}^{\rm{U}}={\rm{Conv}}\Big(\bigcup\nolimits_{q_1\in[x_1,x_2]}{\mathcal{C}}^{\rm{U}}(q_1)\Big).
\end{align}
\end{lemma}
\vspace{-5pt}
\begin{IEEEproof}
See Appendix \ref{Proof_Lemma_Single_PA_Basic_Step} for more details.
\end{IEEEproof}
\vspace{-5pt}
\begin{remark}
The results in Lemma \ref{Lemma_Single_PA_Basic_Step} suggest that the rate pairs on the frontier or Pareto boundary of the uplink capacity region are achieved by setting the activated position along the line segment connecting the projections of the users, i.e., $q_1\in[x_1,x_2]$, which aligns with intuition.
\end{remark}
\vspace{-5pt}
\vspace{-5pt}
\begin{remark}\label{Capacity_Region_Comparision_Uplink}
Recalling that $q_{\rm{f}}\in[q_0,q_{\max}]$, and combining this with \eqref{Single_Pinch_Exhausitive_Search_Capacity} and \eqref{Single_Pinch_Single_PA_Basic_Step}, it follows that ${\mathcal{C}}_{\rm{f}}^{\rm{U}}\subseteq{\mathcal{C}}^{\rm{U}}$, which implies that the uplink capacity region of a conventional fixed-antenna system nests in the capacity region achieved by PASS.
\end{remark}
\vspace{-5pt}
Using Lemma \ref{Lemma_Single_PA_Basic_Step}, we further derive a closed-form expression for the capacity region in this single-pinch case. With $q_1\in[x_1,x_2]$, problem \eqref{Rate_Profile_Uplink_Simplified} reduces to the following:
\begin{subequations}\label{Rate_Profile_Uplink_Single_Pinch}
\begin{align}
\max_{q_1,R}~R~~{\rm{s.t.}}~&q_1\in[x_1,x_2],R\geq0,\\
&f_{{\bm\pi}}^{(1)}(q_1)\geq \overline{\alpha} R,f_{{\bm\pi}}^{(2)}(q_1)\geq \alpha R.
\end{align}
\end{subequations}
where $f_{{\bm\pi}}^{(1)}(x)\triangleq\log_2\left(1+\frac{\frac{\eta P_{\pi_1}}{d_{\pi_1}^2+(x_{\pi_1}-x)^2}}{\frac{\eta P_{\pi_2}}{d_{\pi_2}^2+(x_{\pi_2}-x)^2}+\sigma^2}\right)$ and $f_{{\bm\pi}}^{(2)}(x)\triangleq\log_2\left(1+\frac{\eta P_{\pi_2}/\sigma^2}{d_{\pi_2}^2+(x_{\pi_2}-q_1)^2}\right)$. Although non-convex, this problem admits a closed-form solution as shown below.
\vspace{-5pt}
\begin{theorem}\label{Theorem_PASS_Uplink_Capacity_Region_SP_Closed-Form}
Given $\{\alpha,{\bm\pi}={\bm\pi}^{\rm{\uppercase\expandafter{\romannumeral2}}}\}$, the optimal activated antenna position $q_1$ that solves problem \eqref{Rate_Profile_Uplink_Single_Pinch} is given by
\begin{align}\label{Theorem_PASS_Uplink_Capacity_Region_SP_Closed-Form_Solution}
q_{{\bm\pi}^{\rm{\uppercase\expandafter{\romannumeral2}}}}^{\alpha}=\left\{\begin{array}{ll}
                     x_1 & \frac{1}{\alpha}f_{{\bm\pi}^{\rm{\uppercase\expandafter{\romannumeral2}}}}^{(2)}(x_1)>\frac{1}{1-\alpha}f_{{\bm\pi}^{\rm{\uppercase\expandafter{\romannumeral2}}}}^{(1)}(x_1) \\
                     x_2 & \frac{1}{1-\alpha}f_{{\bm\pi}^{\rm{\uppercase\expandafter{\romannumeral2}}}}^{(1)}(x_2)>\frac{1}{\alpha}f_{{\bm\pi}^{\rm{\uppercase\expandafter{\romannumeral2}}}}^{(2)}(x_2) \\
                     x_{\alpha}^{\star} & {\rm{Else}} 
                   \end{array}\right.,
\end{align}
where $x_{\alpha}^{\star}$ is the solution of $x$ to the equation $\frac{1}{\alpha}f_{{\bm\pi}^{\rm{\uppercase\expandafter{\romannumeral2}}}}^{(2)}(x)
=\frac{1}{1-\alpha}f_{{\bm\pi}^{\rm{\uppercase\expandafter{\romannumeral2}}}}^{(1)}(x)$ with $x\in[x_1,x_2]$.
\end{theorem}
\vspace{-5pt}
\begin{IEEEproof}
See Appendix \ref{Proof_Theorem_PASS_Uplink_Capacity_Region_SP_Closed-Form} for more details.
\end{IEEEproof}
The corresponding solution $q_{{\bm\pi}^{\rm{\uppercase\expandafter{\romannumeral1}}}}^{\alpha}$ under decoding order ${\bm\pi}={\bm\pi}^{\rm{\uppercase\expandafter{\romannumeral1}}}$ can be obtained via a similar approach. Using these results, the complete single-pinch capacity region is given by
\begin{align}\label{Uplink_Channel_Capacity_Region_Rate_Profile_Basic_Single_Pinch}
{\mathcal{C}}^{\rm{U}}={\rm{Conv}}\Big(\bigcup\nolimits_{\alpha\in[0,1],{\bm\pi}\in\{{\bm\pi}^{\rm{\uppercase\expandafter{\romannumeral1}}},
{\bm\pi}^{\rm{\uppercase\expandafter{\romannumeral2}}}\}}{\mathcal{C}}_{{\bm\pi}}^{\rm{U}}({{q}}_{{\bm\pi}}^{\alpha})\Big)\triangleq {\mathcal{C}}_{\rm{S}}^{\rm{U}}.
\end{align}

Next, we derive the achievable rate regions with TDMA and FDMA, where the \emph{SIC/joint decoding} is not required since the signals from the two users are orthogonally separated in the time and frequency domains, respectively. The resulting rate regions constitute inner bounds of the capacity region \cite{el2011network}.
\subsubsection{Achievable Data Rate Region With TDMA}
With TDMA, the two users transmit in two orthogonal time slots. Let $\rho_{\rm{T}}\in[0,1]$ denote the time fraction allocated to user $1$. Since the users are separated in time, the BS can optimize the activated antenna position independently for each time slot. Let $q[k]$ denote the activated position during the time slot assigned to user $k$. For any given $\{q[k]\}_{k\in{\mathcal{K}}}$, the achievable rate region under TDMA is given by ${\mathcal{R}}_{{\rm{S}},{\rm{T}}}^{\rm{U}}(\{q[k]\})\triangleq\bigcup\nolimits_{\rho_{\rm{T}}\in[0,1]}\{(R_1^{\rm{U}},R_2^{\rm{U}}):0\leq R_1^{\rm{U}}\leq\rho_{\rm{T}}\log_2(1+P_1\lvert h_{1}(q[1])\rvert^2/{\sigma^2}),0\leq R_2^{\rm{U}}\leq(1-\rho_{\rm{T}})\log_2(1+P_2\lvert h_{2}(q[2])\rvert^2/{\sigma^2})\}$. Referring to \eqref{Uplink_SU_Channel_Gain_Basic}, the per-user channel power $\lvert h_{k}(q[k])\rvert^2$ can be maximized as $\frac{\eta}{d_k^2}$ by setting $q[k]=x_k$ for $k\in{\mathcal{K}}$. Therefore, the TDMA rate region of the uplink PASS can be written as follows:
\begin{align}
{\mathcal{R}}_{{\rm{S}},{\rm{T}}}^{\rm{U}}\triangleq\bigcup\nolimits_{\rho_{\rm{T}}\in[0,1]}&\{(R_1^{\rm{U}},R_2^{\rm{U}}):0\leq R_1^{\rm{U}}\leq\rho_{\rm{T}}\log_2(1+\gamma_1),\nonumber\\
&0\leq R_2^{\rm{U}}\leq(1-\rho_{\rm{T}})\log_2(1+\gamma_2)\},\label{Uplink_TDMA_Rate_Region_Single_Pinch}
\end{align}
where $\gamma_k\triangleq\frac{P_k\eta}{d_k^2{\sigma^2}}$ for $k\in{\mathcal{K}}$.
\subsubsection{Achievable Data Rate Region With FDMA}\label{Section: Single_Pinch: Achievable Rate Region With FDMA}
With FDMA, the two users transmit simultaneously over orthogonal frequency bands. Let $\rho_{\rm{F}}\in[0,1]$ denote the bandwidth fraction allocated to user $1$. For a given antenna position $q_1$, the FDMA achievable rate region is given by ${\mathcal{R}}_{{\rm{S}},{\rm{F}}}^{\rm{U}}(q_1)\triangleq\bigcup\nolimits_{\rho_{\rm{F}}\in[0,1]}\{(R_1^{\rm{U}},R_2^{\rm{U}}):0\leq R_1^{\rm{U}}\leq r_{1}^{\rho_{\rm{F}}}(q_1),0\leq R_2^{\rm{U}}\leq r_{2}^{\rho_{\rm{F}}}(q_1)\}$, where $r_{1}^{\rho_{\rm{F}}}(x)\triangleq{\rho_{\rm{F}}}\log_2\left(1+\frac{\eta P_{1}/(\rho_{\rm{F}}\sigma^2)}{d_{1}^2+(x-x_1)^2}\right)$ and $r_{2}^{\rho_{\rm{F}}}(x)\triangleq({1-\rho_{\rm{F}}})\log_2\left(1+\frac{\eta P_{2}/((1-\rho_{\rm{F}}){\sigma^2})}{d_{2}^2+(x-x_2)^2}\right)$. By applying \emph{time sharing} among different $q_1$'s and following the proof of Lemma \ref{Lemma_Single_PA_Basic_Step}, the overall FDMA achievable rate region becomes ${\mathcal{R}}_{{\rm{S}},{\rm{F}}}^{\rm{U}}\triangleq{\rm{Conv}}(\bigcup\nolimits_{q_1\in[x_1,x_2]}{\mathcal{R}}_{{\rm{F}}}^{\rm{U}}(q_1))$. This region can also be characterized via the \emph{rate-profile} approach, by solving the following problem for a given rate-profile factor $\alpha_{\rm{F}}\in[0,1]$:
\begin{subequations}\label{Rate_Profile_Uplink_Single_Pinch_FDMA}
\begin{align}
\max_{q_1,R}~R\quad{\rm{s.t.}}~&q_1\in[x_1,x_2],R\geq0,\\
&r_{1}^{\rho_{\rm{F}}}(q_1)\geq \alpha_{\rm{F}} R,r_{2}^{\rho_{\rm{F}}}(q_1)\geq (1-{\alpha}_{\rm{F}}) R.
\end{align}
\end{subequations}
The solution to problem \eqref{Rate_Profile_Uplink_Single_Pinch_FDMA} is given as follows.
\vspace{-5pt}
\begin{theorem}\label{Theorem_PASS_Uplink_Single_Pinch_FDMA}
Given $\alpha_{\rm{F}}$, the optimal activated antenna position $q_1$ that solves problem \eqref{Rate_Profile_Uplink_Single_Pinch_FDMA} is given by
\begin{align}\label{Theorem_PASS_Uplink_Single_Pinch_FDMA_Solution}
q_{\alpha_{\rm{F}}}^{\rho_{\rm{F}}}=\left\{\begin{array}{ll}
                     x_1 & \frac{1}{1-\alpha_{\rm{F}}}r_{2}^{\rho_{\rm{F}}}(x_1)>\frac{1}{\alpha_{\rm{F}}}r_{1}^{\rho_{\rm{F}}}(x_1) \\
                     x_2 & \frac{1}{\alpha_{\rm{F}}}r_{1}^{\rho_{\rm{F}}}(x_2)>\frac{1}{1-\alpha_{\rm{F}}}r_{2}^{\rho_{\rm{F}}}(x_2) \\
                     x_{\alpha_{\rm{F}}}^{\rho_{\rm{F}}} & {\rm{Else}} 
                   \end{array}\right.,
\end{align}
where $x_{\alpha_{\rm{F}}}^{\rho_{\rm{F}}}$ is the solution of $x$ to the equation $\frac{1}{1-\alpha_{\rm{F}}}r_{2}^{\rho_{\rm{F}}}(x)=\frac{1}{\alpha_{\rm{F}}}r_{1}^{\rho_{\rm{F}}}(x)$ with $x\in[x_1,x_2]$.
\end{theorem}
\vspace{-5pt}
\begin{IEEEproof}
Similar to the proof of Theorem \ref{Theorem_PASS_Uplink_Capacity_Region_SP_Closed-Form}.
\end{IEEEproof}
Based on Theorem \ref{Theorem_PASS_Uplink_Single_Pinch_FDMA}, the complete FDMA achievable rate region in the single-pinch case is given by
\begin{align}\label{Uplink_FDMA_Rate_Region_Single_Pinch}
{\mathcal{R}}_{{\rm{S}},{\rm{F}}}^{\rm{U}}={\rm{Conv}}\Big(\bigcup\nolimits_{\alpha_{\rm{F}}\in[0,1],\rho_{\rm{F}}\in[0,1]}{\mathcal{R}}_{\rho_{\rm{F}}}^{\rm{U}}(q_{\alpha_{\rm{F}}}^{\rho_{\rm{F}}})\Big),
\end{align}
where ${\mathcal{R}}_{\rho_{\rm{F}}}^{\rm{U}}(q_1)\triangleq\{(R_1^{\rm{U}},R_2^{\rm{U}}):0\leq R_k^{\rm{U}}\leq r_{k}^{\rho_{\rm{F}}}(q_1),k\in{\mathcal{K}}\}$.
\subsubsection{Comparison}
For any activated position $q_1\in[q_0,q_{\max}]$, the FDMA rate region is always a subset of the corresponding capacity region, i.e., ${\mathcal{R}}_{{\rm{S}},{\rm{F}}}^{\rm{U}}(q_1)\subseteq{\mathcal{C}}^{\rm{U}}(q_1)$ \cite{el2011network}, which implies that ${\mathcal{R}}_{{\rm{S}},{\rm{F}}}^{\rm{U}}\subseteq{\mathcal{C}}_{\rm{S}}^{\rm{U}}$. We next compare the FDMA rate region with the TDMA rate region.
\vspace{-5pt}
\begin{theorem}\label{Theorem_Comparision_TDMA_FDMA}
Given any user locations $\{{\mathbf{u}}_{k}\}_{k\in{\mathcal{K}}}$, the TDMA rate region nests in the FDMA rate region, i.e., ${\mathcal{R}}_{{\rm{S}},{\rm{T}}}^{\rm{U}}\subseteq{\mathcal{R}}_{{\rm{S}},{\rm{F}}}^{\rm{U}}$.
\end{theorem}
\vspace{-5pt}
\begin{IEEEproof}
See Appendix \ref{Proof_Theorem_Comparision_TDMA_FDMA} for more details.
\end{IEEEproof}
\vspace{-5pt}
\begin{remark}
Theorem \ref{Theorem_Comparision_TDMA_FDMA} provides a valuable theoretical foundation for the practical deployment of PASS with frequency-division schemes. As orthogonal FDMA (OFDMA) is expected to remain a key multiple access technology in 6G, this result supports the viability of integrating PASS into both existing and future network architectures, which highlights FDMA as a strong candidate for practical PASS implementations.
\end{remark}
\vspace{-5pt}
Taken together, in the single-pinch case, we conclude
\begin{align}\label{Uplink_Region_Comparison_Single_Pinch}
{\mathcal{R}}_{{\rm{S}},{\rm{T}}}^{\rm{U}}\subseteq{\mathcal{R}}_{{\rm{S}},{\rm{F}}}^{\rm{U}}\subseteq{\mathcal{C}}_{\rm{S}}^{\rm{U}}.
\end{align}
\subsection{Multiple-Pinch Case}\label{Section: Multiple-Pinch Case: Uplink}
We now consider the multiple-pinch case with $N>1$, where the pinching beamforming must be carefully designed. 
\subsubsection{Capacity Region}
Unlike the single-pinch scenario analyzed in \eqref{Rate_Profile_Uplink_Single_Pinch}, the \emph{rate-profile} problem formulated in \eqref{Rate_Profile_Uplink_General} does not admit a closed-form solution when $N>1$ and typically requires an exhaustive search over ${\mathbf{q}}\in{\mathcal{P}}$ to determine the rate region ${\mathcal{C}}_{{\bm\pi}}^{\rm{U}}({\mathbf{q}}_{{\bm\pi}}^{\alpha})$. To address this challenge, we instead aim to develop both an \emph{inner bound} and an \emph{outer bound} on ${\mathcal{C}}^{\rm{U}}$.
\subsubsection*{Capacity Region Inner Bound}
We now derive an inner bound on ${\mathcal{C}}^{\rm{U}}$ based on the rate-profile approach. Specifically, we employ an element-wise alternating optimization framework to efficiently obtain a high-quality suboptimal solution to the non-convex problem in \eqref{Rate_Profile_Uplink_General}.

In the proposed approach, each antenna position $q_n$ is optimized sequentially while keeping all other coordinates fixed. For each $n\in{\mathcal{N}}$, the subproblem becomes
\begin{subequations}\label{MP_Pareto_Sub}
\begin{align}
\max_{q_n,R}~&R\\{\rm{s.t.}}~&\lvert q_n-q_{n'} \rvert\geq\Delta,n\ne n',q_n\in[q_0,q_{\max}],\label{MP_Uplink_Region__Pareto_Sub_Cons1}\\
&{\mathcal{R}}_{1,n}^{\rm{U}}({\bm\pi},q_n,{\mathbf{p}}^{\star})\geq \alpha R,{\mathcal{R}}_{2,n}^{\rm{U}}({\bm\pi},q_n,{\mathbf{p}}^{\star})\geq \overline{\alpha} R,
\end{align}
\end{subequations}
where ${\mathbf{p}}^{\star}\triangleq[P_{\pi_1},P_{\pi_2}]^{\mathsf{T}}$, and ${\mathcal{R}}_{k,n}^{\rm{U}}({\bm\pi},x,{\mathbf{p}}^{\star})$ denotes the achievable rate of the $k$th decoded user under decoding order $\bm\pi$, when $q_n$ in $\mathbf{q}$ is set to $x$ while others are fixed. This subproblem is equivalent to the following:
\begin{align}\label{MP_Pareto_Sub_sub1}
\max_{q_n}{\min_{k\in\{1,2\}}\left\{{\alpha_k^{-1}}{\mathcal{R}}_{k,n}^{\rm{U}}({\bm\pi},q_n,{\mathbf{p}}^{\star})\right\}}~~{\rm{s.t.}}~\eqref{MP_Uplink_Region__Pareto_Sub_Cons1},
\end{align}
where $\alpha_1=\alpha$ and $\alpha_2=1-\alpha$. Since this is a single-variable optimization over a bounded interval, it can be solved efficiently via one-dimensional search. To do so, we discretize the interval $[q_0, q_{\max}]$ into a uniform grid of $Q$ points:
\begin{align}
{\mathcal{Q}}\triangleq \left\{q_0,q_0+\frac{I_{\rm{P}}}{Q-1},q_0+\frac{2I_{\rm{P}}}{Q-1},\ldots,q_{\max}\right\},
\end{align}
where $I_{\rm{P}}\triangleq q_{\max}-q_0$. A near-optimal $q_n$, denoted as $\underline{q}_n^{\star}$, is then selected according to the following:
\begin{align}
\underline{q}_n^{\star}\triangleq\argmax_{q_n\in {\mathcal{Q}}\setminus {\mathcal{Q}}_n}{\min_{k\in{\mathcal{K}}}\left\{{\alpha_k^{-1}}{\mathcal{R}}_{k,n}^{\rm{U}}({\bm\pi},q_n,{\mathbf{p}}^{\star})\right\}},
\end{align}
where the set ${\mathcal{Q}}_n$ includes all grid points that violate the minimum spacing constraint, i.e.,
\begin{align}
{\mathcal{Q}}_n\triangleq \{x|x\in{\mathcal{Q}},\lvert x - q_{n'}\rvert<\Delta,n'\ne n\}.
\end{align}
This procedure is applied iteratively to all antennas until convergence. The complete algorithm is summarized in Algorithm \ref{Algorithm1}, with a computational complexity of ${\mathcal{O}}(I_{\rm{iter}} NQ)$, where $I_{\rm{iter}}$ is the number of iterations to convergence.

\begin{algorithm}[!t]  
\caption{Element-wise Algorithm for Solving \eqref{Rate_Profile_Uplink_General}}
\label{Algorithm1}
\begin{algorithmic}[1]
\STATE initialize the optimization variables
\REPEAT 
  \FOR{$n\in\{1,\ldots,N\}$}
      \STATE update $q_n$ by solving problem \eqref{MP_Pareto_Sub_sub1} through one-dimensional search
    \ENDFOR
\UNTIL{the fractional decrease of the objective value of problem \eqref{MP_Pareto_Sub} falls below a predefined threshold}
\end{algorithmic}
\end{algorithm}

Let ${\underline{\mathbf{q}}}_{{\bm\pi}}^{\alpha}$ denote the optimized pinching beamformer for given $\alpha\in[0,1]$ and ${\bm\pi}\in\{{\bm\pi}^{\rm{\uppercase\expandafter{\romannumeral1}}},
{\bm\pi}^{\rm{\uppercase\expandafter{\romannumeral2}}}\}$. Following the derivation of \eqref{Uplink_Channel_Capacity_Region_Rate_Profile_Basic}, the uplink rate region of multiple-pinch PASS achieved by the element-wise algorithm is characterized as follows:
\begin{align}\label{Uplink_Channel_Capacity_Region_Inner_Bound_Rate_Profile_Basic}
{\mathcal{C}}_{{\rm{M}},{\rm{IB}}}^{\rm{U}}\triangleq{\rm{Conv}}\Big(\bigcup\nolimits_{\alpha\in[0,1],{\bm\pi}\in\{{\bm\pi}^{\rm{\uppercase\expandafter{\romannumeral1}}},
{\bm\pi}^{\rm{\uppercase\expandafter{\romannumeral2}}}\}}{\mathcal{C}}_{{\bm\pi}}^{\rm{U}}({\underline{\mathbf{q}}}_{{\bm\pi}}^{\alpha})\Big),
\end{align}
which satisfies ${\mathcal{C}}_{{\rm{M}},{\rm{IB}}}^{\rm{U}}\subseteq{\mathcal{C}}_{{\rm{M}}}^{\rm{U}}$, where ${\mathcal{C}}_{{\rm{M}}}^{\rm{U}}$ represents the complete uplink capacity region defined in \eqref{Uplink_Channel_Capacity_Region_Rate_Profile_Basic} for $N>1$. 
\subsubsection*{Capacity Region Outer Bound}
We next derive an outer bound on ${\mathcal{C}}_{{\rm{M}}}^{\rm{U}}$. Specifically, it follows from \eqref{PASS_Uplink_Capacity_Region_Constraint1} and \eqref{PASS_Uplink_Capacity_Region_Constraint2} that an outer bound of ${\mathcal{C}}_{{\rm{M}}}^{\rm{U}}$ can be constructed by upper-bounding $r_1^{\rm{U}}({\mathbf{q}})$, $r_2^{\rm{U}}({\mathbf{q}})$, and $r_{1,2}^{\rm{U}}({\mathbf{q}})$, separately.

Referring to \eqref{PASS_Uplink_Capacity_Region_Constraint1}, maximizing $r_k^{\rm{U}}({\mathbf{q}})$ is equivalent to maximizing the effective channel gain $\lvert h_k({\mathbf{q}})\rvert$ of each user $k\in{\mathcal{K}}$, which leads to
\begin{align}\label{Per_User_Channel_Gain_Maximization}
\argmax\nolimits_{{\mathbf{q}}\in{\mathcal{P}}}r_k^{\rm{U}}({\mathbf{q}})=
\argmax\nolimits_{{\mathbf{q}}\in{\mathcal{P}}}\lvert h_k({\mathbf{q}})\rvert^2\triangleq {\mathbf{q}}_k^{\diamond}.
\end{align}
Despite its non-convexity, this problem can be addressed using the antenna position refinement method developed in \cite{xu2024rate}. According to \cite{xu2024rate}, the optimal solution ${\mathbf{q}}_k^{\diamond}$ aims to position the active antennas as close as possible to user $k$, subject to the minimum spacing constraint.

Next, we derive an upper bound for the sum-rate $r_{1,2}^{\rm{U}}({\mathbf{q}})=\log_2(1+(P_1\lvert h_1({\mathbf{q}})\rvert^2
+P_2\lvert h_2({\mathbf{q}})\rvert^2)/\sigma^2)$, where $h_k({\mathbf{q}})=\frac{1}{\sqrt{N}}{\mathbf{h}}^{\mathsf{T}}({\mathbf{u}}_{k},{\mathbf{q}}){\bm\phi}$ for $k\in{\mathcal{K}}$. Relaxing the feasible set ${\mathcal{P}}$ yields
\begin{align}
\max\nolimits_{{\mathbf{q}}\in{\mathcal{P}}}r_{1,2}^{\rm{U}}({\mathbf{q}})\leq \max\nolimits_{q_n\in[q_0,q_{\max}],\forall n\in{\mathcal{N}}}r_{1,2}^{\rm{U}}({\mathbf{q}}).
\end{align}
Applying the \emph{Cauchy–Schwarz inequality}, we obtain
\begin{align}\label{array_gain_upper_bound}
\lvert h_k({\mathbf{q}})\rvert^2&\leq\frac{\lVert{\mathbf{h}}({\mathbf{u}}_{k},{\mathbf{q}})\rVert^2\lVert{\bm\phi}\rVert^2}{N}
=\sum_{n=1}^{N}\frac{\eta}{{d_k^2+(x_k-q_n)^2}}.
\end{align}
Therefore, we can upper bound $\max_{{\mathbf{q}}\in{\mathcal{P}}}r_{1,2}^{\rm{U}}({\mathbf{q}})$ as follows:
\begin{align}\label{Sum_Rate_Capacity_Upper_Bound_Mediate}
\max\nolimits_{{\mathbf{q}}\in{\mathcal{P}}}r_{1,2}^{\rm{U}}({\mathbf{q}})\leq\log_2\left(1+N\max\nolimits_{x\in[q_0,q_{\max}]}f_{\diamond}(x)\right),
\end{align}
where $f_{\diamond}(x)\triangleq\frac{{P_1\eta}/{\sigma^2}}{{d_1^2+(x-x_1)^2}}
+\frac{{P_2\eta}/{\sigma^2}}{{d_2^2+(x-x_2)^2}}$. Moreover, since $x_k\in[x_1,x_2]\subseteq[q_0,q_{\max}]$ for $k\in{\mathcal{K}}$, we have
\begin{align}
\max\nolimits_{x\in[q_0,q_{\max}]}f_{\diamond}(x)=
\max\nolimits_{x\in[x_1,x_2]}f_{\diamond}(x).
\end{align}
Hence, the upper bound becomes
\begin{subequations}
\begin{align}
r_{1,2}^{\rm{U}}({\mathbf{q}})&\leq\max\nolimits_{{\mathbf{q}}\in{\mathcal{P}}}r_{1,2}^{\rm{U}}({\mathbf{q}})\\
&\leq\log_2\left(1+N\max\nolimits_{x\in[x_1,x_2]}f_{\diamond}(x)\right).
\end{align}
\end{subequations}
To evaluate this maximum, we solve
\begin{align}
\argmax\nolimits_{x\in[x_1,x_2]}f_{\diamond}(x)\triangleq x^{\diamond}.
\end{align}

As $f_{\diamond}(x)$ is continuous over the compact interval $[x_1,x_2]$, the maximum occurs either at the endpoints or at a critical point where the derivative vanishes. Taking the derivative:
\begin{align}
\frac{{\rm{d}}f_{\diamond}(x)}{{\rm{d}}x}=\frac{\frac{-2P_1\eta}{\sigma^2}(x-x_1)}{({d_1^2+(x-x_1)^2})^2}
+\frac{\frac{-2P_2\eta}{\sigma^2}(x-x_2)}{({d_2^2+(x-x_2)^2})^2},
\end{align}
and setting $\frac{{\rm{d}}f_{\diamond}(x)}{{\rm{d}}x}=0$, we get the following equation:
\begin{equation}
\begin{split}
&{P_1}(x-x_1)({d_2^2+(x-x_2)^2})^2\\
&+{P_2}(x-x_2)({d_2^2+(x-x_2)^2})^2=0.
\end{split}
\end{equation}
This is a quintic polynomial, which typically does not admit a closed-form solution. However, the solution can be efficiently approximated using Newton's method or grid search. Let ${\mathcal{X}}_{\diamond}$ store the critical points of $f_{\diamond}(x)$ within $x\in[x_1,x_2]$, yielding
\begin{align}
r_{1,2}^{\rm{U}}({\mathbf{q}})\leq\log_2\left(1+N\max\nolimits_{x\in{\mathcal{X}}_{\diamond}\bigcup\{x_1,x_2\}}f_{\diamond}(x)\right)\triangleq
r_{\diamond}^{\rm{U}}.
\end{align} 

Putting all bounds together, the outer bound on the uplink capacity region for $N>1$ can be written as follows:
\begin{align}
{\mathcal{C}}_{{\rm{M}},{\rm{OB}}}^{\rm{U}}\triangleq\left\{(R_1^{\rm{U}},R_2^{\rm{U}})\left\lvert 
\begin{matrix}R_k^{\rm{U}}\in[0,r_k^{\rm{U}}({\mathbf{q}}_k^{\diamond})],\forall k,\\
R_1^{\rm{U}}+R_2^{\rm{U}}\leq r_{\diamond}^{\rm{U}}\end{matrix}\right.\right\}\supseteq{\mathcal{C}}_{{\rm{M}}}^{\rm{U}}.
\end{align}

In summary, we conclude that ${\mathcal{C}}_{{\rm{M}},{\rm{IB}}}^{\rm{U}}\subseteq{\mathcal{C}}_{{\rm{M}}}^{\rm{U}}\subseteq{\mathcal{C}}_{{\rm{M}},{\rm{OB}}}^{\rm{U}}$. Notably, the inner bound becomes exact when $N=1$, as the problem formulated in \eqref{Rate_Profile_Uplink_General} can be optimally solved in this case. Furthermore, the outer bound also remains valid as a bound on the capacity region for $N=1$. The tightness of these two bounds will be further evaluated through numerical results presented in Section \ref{Section_Numerical_Results}. We next characterize the achievable rate regions with TDMA and FDMA.

\begin{table*}[!t]
\centering
\caption{Summary of Main Results for PASS-Enabled Uplink Channels}
\setlength{\abovecaptionskip}{0pt}
\resizebox{0.8\textwidth}{!}{
\begin{tabular}{|l|lll|}
\hline
               & \multicolumn{1}{l|}{Capacity Region}                                                                                                                                        & \multicolumn{1}{l|}{TDMA Rate Region}                                                                                                       & FDMA Rate Region                                                                                                                                       \\ \hline
Single-Pinch   & \multicolumn{1}{l|}{${\mathcal{C}}_{\rm{S}}^{\rm{U}}$ (\emph{closed-form}, \eqref{Uplink_Channel_Capacity_Region_Rate_Profile_Basic_Single_Pinch})}                                                                                & \multicolumn{1}{l|}{${\mathcal{R}}_{{\rm{S}},{\rm{T}}}^{\rm{U}}$ (\emph{closed-form}, \eqref{Uplink_TDMA_Rate_Region_Single_Pinch})}                                                & ${\mathcal{R}}_{{\rm{S}},{\rm{F}}}^{\rm{U}}$ (\emph{closed-form}, \eqref{Uplink_FDMA_Rate_Region_Single_Pinch})                                                                              \\ \hline
Multiple-Pinch & \multicolumn{1}{l|}{\begin{tabular}[c]{@{}l@{}}${\mathcal{C}}_{{\rm{M}},{\rm{IB}}}^{\rm{U}}$ (\emph{rate-profile}); ${\mathcal{C}}_{{\rm{M}},{\rm{OB}}}^{\rm{U}}$ (\emph{relaxation})\\ ${\mathcal{C}}_{{\rm{M}},{\rm{IB}}}^{\rm{U}}\subseteq{\mathcal{C}}_{{\rm{M}}}^{\rm{U}}\subseteq{\mathcal{C}}_{{\rm{M}},{\rm{OB}}}^{\rm{U}}$\end{tabular}} & \multicolumn{1}{l|}{${\mathcal{R}}_{{\rm{M}},{\rm{T}}}^{\rm{U}}$ (\emph{method in \cite{xu2024rate}})} & \begin{tabular}[c]{@{}l@{}}${\mathcal{C}}_{{\rm{M}},{\rm{IB}}}^{\rm{U}}$ (\emph{rate-profile}); ${\mathcal{C}}_{{\rm{M}},{\rm{OB}}}^{\rm{U}}$ (\emph{relaxation})\\ ${\mathcal{R}}_{{\rm{M}},{\rm{F}},{\rm{IB}}}^{\rm{U}}\subseteq{\mathcal{R}}_{{\rm{M}},{\rm{F}}}^{\rm{U}}\subseteq{\mathcal{C}}_{{\rm{M}},{\rm{OB}}}^{\rm{U}}$\end{tabular} \\ \hline
Comparison    & \multicolumn{3}{l|}{${\mathcal{R}}_{{\rm{S}},{\rm{T}}}^{\rm{U}}\subseteq{\mathcal{R}}_{{\rm{S}},{\rm{F}}}^{\rm{U}}\subseteq{\mathcal{C}}_{\rm{S}}^{\rm{U}}$ (\emph{single-pinch}); ${\mathcal{R}}_{{\rm{M}},{\rm{T}}}^{\rm{U}}\subseteq{\mathcal{R}}_{{\rm{M}},{\rm{F}}}^{\rm{U}}\subseteq{\mathcal{C}}_{\rm{M}}^{\rm{U}}\subseteq{\mathcal{C}}_{{\rm{M}},{\rm{OB}}}^{\rm{U}}$ (\emph{multiple-pinch})}                                                                                                                                                                                                                                                                                                                                                                                                                                                            \\ \hline
\end{tabular}}
\label{Table: PASS_Uplink_Capacity_Region}
\vspace{-10pt}
\end{table*}

\subsubsection{Achievable Data Rate Region With TDMA}
Under the TDMA scheme, the two users are allocated orthogonal time slots. The pinching beamformer is independently optimized for each time slot. According to the optimal antenna positioning in \eqref{Per_User_Channel_Gain_Maximization}, during the time slot allocated to user $k$, the pinching beamformer is set to ${\mathbf{q}}_k^{\diamond}$. Therefore, the corresponding TDMA achievable rate region is given by
\begin{equation}\label{TDMA_Uplink_Multiple_Pinch_Region_Expression}
\begin{split}
{\mathcal{R}}_{{\rm{M}},{\rm{T}}}^{\rm{U}}\triangleq\bigcup\nolimits_{\rho_{\rm{T}}\in[0,1]}&\{(R_1^{\rm{U}},R_2^{\rm{U}}):0\leq R_1^{\rm{U}}\leq\rho_{\rm{T}}r_1^{\rm{U}}({\mathbf{q}}_1^{\diamond}),\\
&0\leq R_2^{\rm{U}}\leq(1-\rho_{\rm{T}})r_2^{\rm{U}}({\mathbf{q}}_2^{\diamond})\},
\end{split}
\end{equation}
where $r_k^{\rm{U}}({\mathbf{q}}_k^{\diamond})=\log_2( P_k\lvert h_k({\mathbf{q}}_k^{\diamond})\rvert^2/\sigma^2)$ for $k\in{\mathcal{K}}$.
\subsubsection{Achievable Data Rate Region With FDMA}\label{Section:Uplink:Multiple_Pinch:Achievable Rate Region With FDMA}
Under FDMA, both users transmit simultaneously over orthogonal frequency bands. For any given $\mathbf{q}$, the FDMA achievable region is given by ${\mathcal{R}}_{{\rm{M}},{\rm{F}}}^{\rm{U}}(\mathbf{q})\triangleq\bigcup\nolimits_{\rho_{\rm{F}}\in[0,1]}\{(R_1^{\rm{U}},R_2^{\rm{U}}):0\leq R_k^{\rm{U}}\leq R_{k}^{\rho_{\rm{F}}}(\mathbf{q}),\forall k\in{\mathcal{K}}\}$, where $R_{1}^{\rho_{\rm{F}}}(\mathbf{q})\triangleq\rho_{\rm{F}}\log_2\left(1+\frac{P_1\lvert h_{1}(\mathbf{q})\rvert^2}{\rho_{\rm{F}}{\sigma^2}}\right)$ and $R_{2}^{\rho_{\rm{F}}}(\mathbf{q})\triangleq({1-\rho_{\rm{F}}})\log_2\left(1+\frac{P_2\lvert h_{2}(\mathbf{q})\rvert^2}{(1-\rho_{\rm{F}}){\sigma^2}}\right)$. After \emph{time sharing} among different $\mathbf{q}$'s, the overall FDMA achievable rate region becomes ${\mathcal{R}}_{{\rm{M}},{\rm{F}}}^{\rm{U}}\triangleq{\rm{Conv}}\left(\bigcup\nolimits_{{\mathbf{q}}\in{\mathcal{P}}}{\mathcal{R}}_{{\rm{M}},{\rm{F}}}^{\rm{U}}(\mathbf{q})\right)$. This region can be characterized via a \emph{rate-profile} approach, similar to the derivation of ${\mathcal{R}}_{{\rm{S}},{\rm{F}}}^{\rm{U}}$ in Section \ref{Section: Single_Pinch: Achievable Rate Region With FDMA}. Given a rate-profile factor $\alpha_{\rm{F}}\in[0,1]$, we solve the following problem:
\begin{subequations}\label{Rate_Profile_Uplink_Multiple_Pinch_FDMA}
\begin{align}
\max_{{\mathbf{q}},R}~R~{\rm{s.t.}}~&{\mathbf{q}}\in{\mathcal{P}},R\geq0,\\
&R_{1}^{\rho_{\rm{F}}}(\mathbf{q})\geq \alpha_{\rm{F}} R,R_{2}^{\rho_{\rm{F}}}(\mathbf{q})\geq (1-{\alpha}_{\rm{F}}) R.
\end{align}
\end{subequations}
A suboptimal solution to this problem can be found using an element-wise alternating optimization method, similar to that used for problem \eqref{Rate_Profile_Uplink_General}. For brevity, the details are omitted. Based on this, an inner bound of ${\mathcal{R}}_{{\rm{M}},{\rm{F}}}^{\rm{U}}$, denoted as ${\mathcal{R}}_{{\rm{M}},{\rm{F}},{\rm{IB}}}^{\rm{U}}$, can be similarly obtained as ${\mathcal{C}}_{{\rm{M}},{\rm{IB}}}^{\rm{U}}$ (see \eqref{Uplink_Channel_Capacity_Region_Inner_Bound_Rate_Profile_Basic}). 
\subsubsection{Comparison and Further Discussion}
For any given ${\mathbf{q}}$, it holds that ${\mathcal{R}}_{{\rm{M}},{\rm{F}}}^{\rm{U}}(\mathbf{q})\subseteq{\mathcal{C}}^{\rm{U}}(\mathbf{q})$ \cite{el2011network}, which implies that ${\mathcal{R}}_{{\rm{M}},{\rm{F}}}^{\rm{U}}\subseteq{\mathcal{C}}_{\rm{M}}^{\rm{U}}$. Following a similar approach to the proof of Theorem \ref{Theorem_Comparision_TDMA_FDMA}, it can be shown that, under the multiple-pinch case, the TDMA region is nested in the FDMA region, i.e., ${\mathcal{R}}_{{\rm{M}},{\rm{T}}}^{\rm{U}}\subseteq{\mathcal{R}}_{{\rm{M}},{\rm{F}}}^{\rm{U}}$. Hence, the following inclusion chain holds: 
\begin{align}\label{Region_Uplink_Multiple_Relationship}
{\mathcal{R}}_{{\rm{M}},{\rm{T}}}^{\rm{U}}\subseteq{\mathcal{R}}_{{\rm{M}},{\rm{F}}}^{\rm{U}}\subseteq{\mathcal{C}}_{\rm{M}}^{\rm{U}}\subseteq{\mathcal{C}}_{{\rm{M}},{\rm{OB}}}^{\rm{U}}.
\end{align}

For ease of reference, we summarize in Table \ref{Table: PASS_Uplink_Capacity_Region} our main results on the uplink capacity/rate region characterization.

We now analyze how the number of activated pinching antennas, $N$, influences the capacity region of the PASS-enabled uplink channel. While it may seem intuitive that increasing 
$N$ should monotonically enlarge the capacity region---owing to enhanced spatial degrees of freedom---this is not always the case, as demonstrated below.

Let us examine the effective array gain of user $k$, which is expressed as follows:
\begin{subequations}\label{Array_Gain_Upper_Bound_Maximization_Pre}
\begin{align}
{\lvert h_k({\mathbf{q}})\rvert^2}&=\frac{\eta}{N}\left\lvert\sum\nolimits_{n=1}^{N}\frac{{\rm{e}}^{-{\rm{j}}\phi_{n}}}{\sqrt{d_k^2+(q_n-x_k)^2}}\right\rvert^2\\
&\leq\frac{\eta}{N}\left\lvert\sum\nolimits_{n=1}^{N}\frac{1}{\sqrt{d_k^2+(q_n-x_k)^2}}\right\rvert^2.
\end{align}
\end{subequations}
It follows that
\begin{align}\label{Array_Gain_Upper_Bound_Maximization}
{\lvert h_k({\mathbf{q}}_k^{\diamond})\rvert^2}\leq\max_{{\mathbf{q}}\in{\mathcal{P}}}\frac{\eta}{N}\left\lvert\sum_{n=1}^{N}\frac{1}{\sqrt{d_k^2+(q_n-x_k)^2}}\right\rvert^2
\triangleq A_{k}^{\triangleleft},
\end{align}
where ${\mathbf{q}}_k^{\diamond}=\argmax\nolimits_{{\mathbf{q}}\in{\mathcal{P}}}\lvert h_k({\mathbf{q}})\rvert^2$ as per \eqref{Per_User_Channel_Gain_Maximization}. The behavior of this upper bound $A_{k}^{\triangleleft}$ as a function of $N$ is characterized by the following lemma.
\vspace{-5pt}
\begin{lemma}\label{Lemma_Array_Gain_Upper_Limitation}
Given a fixed user position ${\mathbf{u}}_k$ and and a minimum inter-antenna spacing $\Delta$, the upper bound on the array gain scales as $A_{k}^{\triangleleft}\simeq{\mathcal{O}}\left({(\ln{N})^2}/{N}\right)$ when $N\rightarrow\infty$.
\end{lemma}
\vspace{-5pt}
\begin{IEEEproof}
See Appendix \ref{Proof_Lemma_Array_Gain_Upper_Limitation} for more details.
\end{IEEEproof}
The result in Lemma \ref{Lemma_Array_Gain_Upper_Limitation} implies that 
\begin{align}
\lim\nolimits_{N\rightarrow\infty}A_{k}^{\triangleleft}=\lim\nolimits_{N\rightarrow\infty}{\mathcal{O}}\left({(\ln{N})^2}/{N}\right)=0,
\end{align}
Combining this with the bounds in \eqref{Array_Gain_Upper_Bound_Maximization_Pre} and \eqref{Array_Gain_Upper_Bound_Maximization} and applying the \emph{Sandwich Theorem}, we obtain
\begin{align}
0\leq\lim_{N\rightarrow\infty}{\lvert h_k({\mathbf{q}})\rvert^2}
\leq\lim_{N\rightarrow\infty}{\lvert h_k({\mathbf{q}}_k^{\diamond})\rvert^2}\leq
\lim_{N\rightarrow\infty}A_{k}^{\triangleleft}=0.
\end{align}
Substituting this result into the individual and sum-rate expressions in \eqref{PASS_Uplink_Capacity_Region_Constraint1} and \eqref{PASS_Uplink_Capacity_Region_Constraint2} gives
\begin{subequations}
\begin{align}
\lim_{N\rightarrow\infty}r_k^{\rm{U}}({\mathbf{q}})&\leq \lim_{N\rightarrow\infty}r_k^{\rm{U}}({\mathbf{q}}_k^{\diamond})=0,\forall k\in{\mathcal{K}},\label{per_user_rate_bound}\\
\lim_{N\rightarrow\infty}r_{1,2}^{\rm{U}}({\mathbf{q}})&\leq\lim_{N\rightarrow\infty}\log_2(1+(P_1A_{1}^{\triangleleft}
+P_2A_{2}^{\triangleleft})/\sigma^2)\\&=\log_2(1+0/\sigma^2)=0.\label{sum_rate_upper_bound}
\end{align}
\end{subequations}
Therefore, the capacity region of any fixed beamformer ${\mathbf{q}}\in{\mathcal{P}}$ satisfies $\lim_{N\rightarrow\infty}{\mathcal{C}}^{\rm{U}}({\mathbf{q}})=\{(0,0)\}$.

Recalling the full capacity region is defined by the convex hull over all feasible beamformers (see \eqref{PASS_Uplink_Capacity_Region_General_Exhausitive}), it follows that
\begin{align}
\lim_{N\rightarrow\infty}{\mathcal{C}}_{\rm{M}}^{\rm{U}}={\rm{Conv}}\Big(\bigcup\nolimits_{{\mathbf{q}}\in{\mathcal{P}}}\{(0,0)\}\Big)=\{(0,0)\},
\end{align}
which, together with \eqref{Region_Uplink_Multiple_Relationship}, yields
\begin{align}\label{Asymptotic_Rate_Capacity_Region_Uplink_Multiple}
\lim_{N\rightarrow\infty}{\mathcal{R}}_{{\rm{M}},{\rm{T}}}^{\rm{U}}=\lim_{N\rightarrow\infty}{\mathcal{R}}_{{\rm{M}},{\rm{F}}}^{\rm{U}}
=\lim_{N\rightarrow\infty}{\mathcal{C}}_{\rm{M}}^{\rm{U}}=\{(0,0)\}.
\end{align}
The results in \eqref{Asymptotic_Rate_Capacity_Region_Uplink_Multiple} can be interpreted as follows. 
\vspace{-5pt}
\begin{remark}\label{remark_uplink_interpretation}
As $N$ increases, the aggregate noise power increases \emph{linearly} as $N\sigma^2$, while the total received signal power grows only \emph{sub-linearly} due to increasing average distance and severe path loss associated with farther positioned pinching elements. Thus, the per-user effective SNR degrades to zero, rendering both individual and sum-rate expressions negligible.
\end{remark}
\vspace{-5pt}

\begin{table*}[!t]
\centering
\caption{Summary of Main Results for PASS-Enabled Downlink Channels}
\setlength{\abovecaptionskip}{0pt}
\resizebox{0.8\textwidth}{!}{
\begin{tabular}{|l|lll|}
\hline
               & \multicolumn{1}{l|}{Capacity Region}                                                                                                                                        & \multicolumn{1}{l|}{TDMA Rate Region}                                                                                                       & FDMA Rate Region                                                                                                                                       \\ \hline
Single-Pinch   & \multicolumn{1}{l|}{${\mathcal{C}}_{\rm{S}}^{\rm{D}}$ (\emph{duality}, \emph{closed-form})}                                                                                & \multicolumn{1}{l|}{${\mathcal{R}}_{{\rm{S}},{\rm{T}}}^{\rm{D}}$ (\emph{duality}, \emph{closed-form})}                                                & ${\mathcal{R}}_{{\rm{S}},{\rm{F}}}^{\rm{D}}$ (\emph{duality}, \emph{closed-form})                                                                              \\ \hline
Multiple-Pinch & \multicolumn{1}{l|}{\begin{tabular}[c]{@{}l@{}}${\mathcal{C}}_{{\rm{M}},{\rm{IB}}}^{\rm{D}}$ (\emph{duality}); ${\mathcal{C}}_{{\rm{M}},{\rm{OB}}}^{\rm{D}}$ (\emph{duality})\\ ${\mathcal{C}}_{{\rm{M}},{\rm{IB}}}^{\rm{D}}\subseteq{\mathcal{C}}_{{\rm{M}}}^{\rm{D}}\subseteq{\mathcal{C}}_{{\rm{M}},{\rm{OB}}}^{\rm{D}}$\end{tabular}} & \multicolumn{1}{l|}{${\mathcal{R}}_{{\rm{M}},{\rm{T}}}^{\rm{D}}$ (\emph{duality}, \emph{method in \cite{xu2024rate}})} & \begin{tabular}[c]{@{}l@{}}${\mathcal{C}}_{{\rm{M}},{\rm{IB}}}^{\rm{D}}$ (\emph{duality}); ${\mathcal{C}}_{{\rm{M}},{\rm{OB}}}^{\rm{D}}$ (\emph{duality})\\ ${\mathcal{R}}_{{\rm{M}},{\rm{F}},{\rm{IB}}}^{\rm{U}}\subseteq{\mathcal{R}}_{{\rm{M}},{\rm{F}}}^{\rm{U}}\subseteq{\mathcal{C}}_{{\rm{M}},{\rm{OB}}}^{\rm{U}}$\end{tabular} \\ \hline
Comparison    & \multicolumn{3}{l|}{${\mathcal{R}}_{{\rm{S}},{\rm{T}}}^{\rm{D}}\subseteq{\mathcal{R}}_{{\rm{S}},{\rm{F}}}^{\rm{D}}\subseteq{\mathcal{C}}_{\rm{S}}^{\rm{D}}$ (\emph{single-pinch}); ${\mathcal{R}}_{{\rm{M}},{\rm{T}}}^{\rm{D}}\subseteq{\mathcal{R}}_{{\rm{M}},{\rm{F}}}^{\rm{D}}\subseteq{\mathcal{C}}_{\rm{M}}^{\rm{D}}\subseteq{\mathcal{C}}_{{\rm{M}},{\rm{OB}}}^{\rm{D}}$ (\emph{multiple-pinch})}                                                                                                                                                                                                                                                                                                                                                                                                                                                            \\ \hline
\end{tabular}}
\label{Table: PASS_Downlink_Capacity_Region}
\vspace{-10pt}
\end{table*}

\section{PASS-Enabled Downlink Channels}\label{Section: PASS-Enabled Downlink Channels}
In this section, we extend the capacity and achievable rate region analysis of the uplink PASS to the downlink PASS by leveraging the \emph{uplink-downlink duality} framework \cite{el2011network,jindal2004duality}.
\subsection{Overview of the Capacity Region}
As a baseline, we first consider the conventional fixed-antenna system described in \eqref{CASS_Downlink_Model}. The capacity region of a two-user downlink channel is achieved using \emph{dirty paper coding (DPC)}, which allows the BS to pre-cancel interference \cite{el2011network}. In our considered case, the capacity region can also be achieved by employing superposition coding combined with SIC decoding at the users, i.e., by utilizing \emph{power-domain non-orthogonal multiple access (NOMA)} \cite{el2011network}. Based on \emph{uplink-downlink duality}, the capacity region of the downlink channel is equivalent to the union of the capacity regions of its dual uplink channels. This equivalence holds under identical effective channels $(h_1^{\rm{f}},h_2^{\rm{f}})$, the same noise power $\sigma^2$, and all power allocations $(P_1,P_2)$ that satisfy $P_1+P_2=P$. Formally, the downlink capacity region is expressed as follows \cite{jindal2004duality}:
\begin{align}\label{Conventional_Fixed_Antenna_Downlink}
{\mathcal{C}}_{\rm{f}}^{\rm{D}}\triangleq\bigcup\nolimits_{P_1+P_2=P}{\mathcal{C}}_{\rm{f}}^{\rm{U}},
\end{align}
where ${\mathcal{C}}_{\rm{f}}^{\rm{U}}$ (see \eqref{Conventional_Fixed_Antenna_Uplink}) denotes the capacity region of the dual uplink channel for a given pair of transmit powers $(P_1,P_2)$. 

\begin{figure}[!t]
\centering
\includegraphics[width=0.3\textwidth]{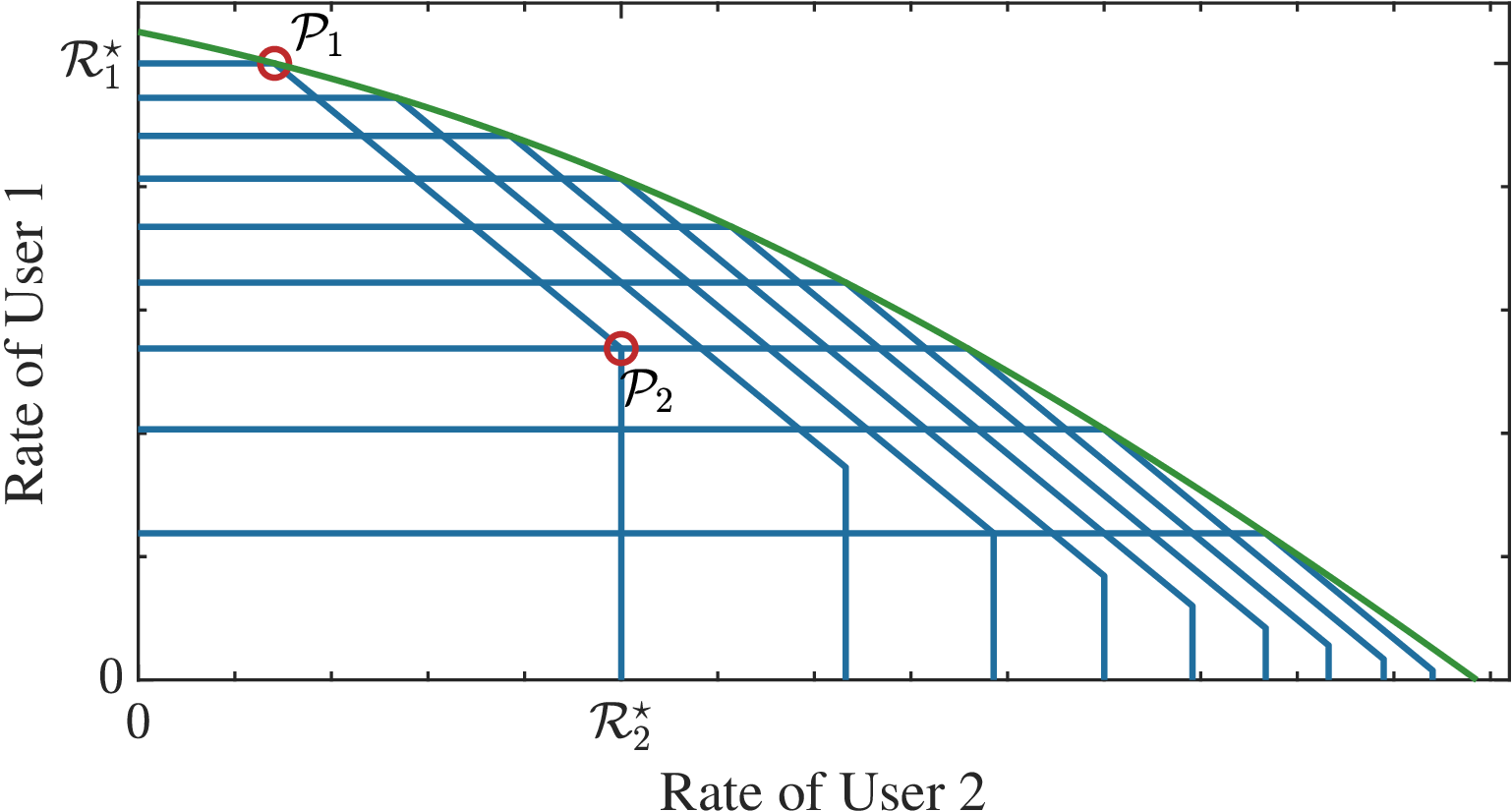}
\caption{Illustrated example of the capacity region for a two-user downlink channel.}
\label{Figure_BC_Capacity_Region_AWGN}
\vspace{-15pt}
\end{figure}

As illustrated in {\figurename} {\ref{Figure_BC_Capacity_Region_AWGN}}, the frontier of the downlink capacity region is formed by connecting the corner points of the associated dual uplink capacity regions. In the figure, the corner points of a dual uplink channel are marked by red circles labeled as ${\mathcal{P}}_1$ and ${\mathcal{P}}_2$. Each blue pentagon represents a dual uplink capacity region associated with a particular power allocation satisfying $P_1+P_2=P$. The green line traces the overall downlink capacity region.

Building upon \eqref{Conventional_Fixed_Antenna_Downlink}, we now characterize the capacity region of the downlink PASS. For any given pinching beamformer ${\mathbf{q}}\in{\mathcal{P}}$, the downlink capacity region is defined as follows:
\begin{align}
{\mathcal{C}}^{\rm{D}}({\mathbf{q}})\triangleq\bigcup\nolimits_{P_1+P_2=P}{\mathcal{C}}^{\rm{U}}({\mathbf{q}}),
\end{align}
where ${\mathcal{C}}^{\rm{U}}({\mathbf{q}})$ denotes the dual uplink capacity region as per \eqref{PASS_Uplink_Capacity_Region_Specific}. By considering \emph{time sharing} among different ${\mathbf{q}}\in{\mathcal{P}}$, the overall downlink capacity region of the PASS is given by
\begin{subequations}\label{Downlink_Capacity_Region_General}
\begin{align}
{\mathcal{C}}^{\rm{D}}&\triangleq{\rm{Conv}}\Big(\bigcup\nolimits_{{\mathbf{q}}\in{\mathcal{P}}}{\mathcal{C}}^{\rm{D}}({\mathbf{q}})\Big)\\
&={\rm{Conv}}\Big(\bigcup\nolimits_{{\mathbf{q}}\in{\mathcal{P}}}\bigcup\nolimits_{P_1+P_2=P}{\mathcal{C}}^{\rm{U}}({\mathbf{q}})\Big)\\
&={\rm{Conv}}\Big(\bigcup\nolimits_{P_1+P_2=P}{\rm{Conv}}\Big(\bigcup\nolimits_{{\mathbf{q}}\in{\mathcal{P}}}{\mathcal{C}}^{\rm{U}}({\mathbf{q}})\Big)\Big)\\
&={\rm{Conv}}\Big(\bigcup\nolimits_{P_1+P_2=P}{\mathcal{C}}^{\rm{U}}\Big).
\end{align}
\end{subequations}
To gain deeper insights into ${\mathcal{C}}^{\rm{D}}$, we next analyze the downlink capacity region and compare it with the achievable rate regions under TDMA and FDMA schemes. For clarity, we begin by considering the single-pinch case.
\subsection{Single-Pinch Case}\label{Section:PASS-Enabled Downlink Channels:Single-Pinch Case}
Inserting \eqref{Uplink_Channel_Capacity_Region_Rate_Profile_Basic_Single_Pinch} into \eqref{Downlink_Capacity_Region_General} gives the downlink capacity region for the single-pinch case as follows:
\begin{align}
{\mathcal{C}}_{\rm{S}}^{\rm{D}}\triangleq{\rm{Conv}}\Big(\bigcup\nolimits_{P_1+P_2=P}{\mathcal{C}}_{\rm{S}}^{\rm{U}}\Big),
\end{align}
which can be computed using the approach detailed in Section \ref{Section: Uplink: Single-Pinch Case: Capacity Region}. Similarly, the achievable downlink rate regions under TDMA and FDMA are derived from their corresponding uplink counterparts ${\mathcal{R}}_{{\rm{S}},{\rm{T}}}^{\rm{U}}$ and ${\mathcal{R}}_{{\rm{S}},{\rm{F}}}^{\rm{U}}$, respectively, which are given by ${\mathcal{R}}_{{\rm{S}},{\rm{T}}}^{\rm{D}}\triangleq{\rm{Conv}}\left(\bigcup\nolimits_{P_1+P_2=P}{\mathcal{R}}_{{\rm{S}},{\rm{T}}}^{\rm{U}}\right)$ and ${\mathcal{R}}_{{\rm{S}},{\rm{F}}}^{\rm{D}}\triangleq{\rm{Conv}}\left(\bigcup\nolimits_{P_1+P_2=P}{\mathcal{R}}_{{\rm{S}},{\rm{F}}}^{\rm{U}}\right)$, respectively. Since ${\mathcal{R}}_{{\rm{S}},{\rm{T}}}^{\rm{U}}\subseteq{\mathcal{R}}_{{\rm{S}},{\rm{F}}}^{\rm{U}}\subseteq{\mathcal{C}}_{\rm{S}}^{\rm{U}}$ holds for the uplink PASS (as per \eqref{Uplink_Region_Comparison_Single_Pinch}), we have ${\mathcal{R}}_{{\rm{S}},{\rm{T}}}^{\rm{D}}\subseteq{\mathcal{R}}_{{\rm{S}},{\rm{F}}}^{\rm{D}}\subseteq{\mathcal{C}}_{\rm{S}}^{\rm{D}}$ for the downlink PASS.
\subsection{Multiple-Pinch Case}
\subsubsection{Capacity Region}
Consider the multiple-pinch case, where the downlink capacity region is defined as follows:
\begin{align}
{\mathcal{C}}_{\rm{M}}^{\rm{D}}\triangleq{\rm{Conv}}\Big(\bigcup\nolimits_{P_1+P_2=P}{\mathcal{C}}_{\rm{M}}^{\rm{U}}\Big),
\end{align}
where ${\mathcal{C}}_{{\rm{M}}}^{\rm{U}}$ represents the complete uplink capacity region defined in \eqref{Uplink_Channel_Capacity_Region_Rate_Profile_Basic} for $N>1$. Since there is no closed-form expression for ${\mathcal{C}}_{{\rm{M}}}^{\rm{U}}$, we establish inner and outer bounds on ${\mathcal{C}}_{\rm{M}}^{\rm{D}}$ based on corresponding bounds for ${\mathcal{C}}_{{\rm{M}}}^{\rm{U}}$. Recalling ${\mathcal{C}}_{{\rm{M}},{\rm{IB}}}^{\rm{U}}\subseteq{\mathcal{C}}_{{\rm{M}}}^{\rm{U}}\subseteq{\mathcal{C}}_{{\rm{M}},{\rm{OB}}}^{\rm{U}}$ yields ${\mathcal{C}}_{{\rm{M}},{\rm{IB}}}^{\rm{D}}\subseteq{\mathcal{C}}_{{\rm{M}}}^{\rm{D}}\subseteq{\mathcal{C}}_{{\rm{M}},{\rm{OB}}}^{\rm{D}}$, where
\begin{align}
{\mathcal{C}}_{{\rm{M}},\clubsuit}^{\rm{D}}\triangleq{\rm{Conv}}\Big(\bigcup\nolimits_{P_1+P_2=P}{\mathcal{C}}_{{\rm{M}},\clubsuit}^{\rm{M}}\Big),\quad \clubsuit\in\{{\rm{OB}},{\rm{IB}}\}.
\end{align}
These bounds are obtained by constructing the dual uplink regions for all combinations $(P_1, P_2)$ satisfying $P_1 + P_2 = P$ and taking the convex hull of the resulting union.
\subsubsection{Achievable Data Rate Region With TDMA}
In the multiple-pinch case, the achievable downlink rate region using TDMA is expressed as ${\mathcal{R}}_{{\rm{M}},{\rm{T}}}^{\rm{D}}\triangleq{\rm{Conv}}\left(\bigcup\nolimits_{P_1+P_2=P}{\mathcal{R}}_{{\rm{M}},{\rm{T}}}^{\rm{U}}\right)$, by referring to its uplink counterpart ${\mathcal{R}}_{{\rm{M}},{\rm{T}}}^{\rm{U}}$, defined in \eqref{TDMA_Uplink_Multiple_Pinch_Region_Expression}, analogous to the single-pinch case.
\subsubsection{Achievable Data Rate Region With FDMA}
For FDMA, we derive inner and outer bounds for the achievable rate region ${\mathcal{R}}_{{\rm{M}},{\rm{F}}}^{\rm{D}}$, in a manner similar to that used for the uplink PASS. The inner bound is constructed using the uplink counterpart ${\mathcal{R}}_{{\rm{M}},{\rm{F}},{\rm{IB}}}^{\rm{U}}$ (see Section \ref{Section:Uplink:Multiple_Pinch:Achievable Rate Region With FDMA}) as ${\mathcal{R}}_{{\rm{M}},{\rm{F}},{\rm{IB}}}^{\rm{D}}\triangleq{\rm{Conv}}\left(\bigcup\nolimits_{P_1+P_2=P}{\mathcal{R}}_{{\rm{M}},{\rm{F}},{\rm{IB}}}^{\rm{U}}\right)$. As for the outer bound, we obtain ${\mathcal{R}}_{{\rm{M}},{\rm{F}}}^{\rm{D}}\subseteq{\mathcal{C}}_{{\rm{M}},{\rm{OB}}}^{\rm{D}}$, due to the fact that ${\mathcal{R}}_{{\rm{M}},{\rm{F}}}^{\rm{D}}\subseteq{\mathcal{C}}_{{\rm{M}}}^{\rm{D}}$ \cite{el2011network}. 
\subsubsection{Comparison and Further Discussion}\label{Section:Downlink PASS:Multiple Pinch:Comparison and Further Discussion}
Since ${\mathcal{R}}_{{\rm{M}},{\rm{T}}}^{\rm{U}}\subseteq{\mathcal{R}}_{{\rm{M}},{\rm{F}}}^{\rm{U}}\subseteq{\mathcal{C}}_{\rm{M}}^{\rm{U}}\subseteq{\mathcal{C}}_{{\rm{M}},{\rm{OB}}}^{\rm{U}}$ holds for the uplink PASS (as per \eqref{Region_Uplink_Multiple_Relationship}), we have ${\mathcal{R}}_{{\rm{M}},{\rm{T}}}^{\rm{D}}\subseteq{\mathcal{R}}_{{\rm{M}},{\rm{F}}}^{\rm{D}}\subseteq{\mathcal{C}}_{\rm{M}}^{\rm{D}}\subseteq{\mathcal{C}}_{{\rm{M}},{\rm{OB}}}^{\rm{D}}$ for the downlink PASS. Furthermore, based on uplink-downlink duality and the uplink asymptotic analyses presented in \eqref{Asymptotic_Rate_Capacity_Region_Uplink_Multiple}, we have
\begin{align}\label{Asymptotic_Rate_Capacity_Region_Downlink_Multiple}
\lim_{N\rightarrow\infty}{\mathcal{R}}_{{\rm{M}},{\rm{T}}}^{\rm{D}}=\lim_{N\rightarrow\infty}{\mathcal{R}}_{{\rm{M}},{\rm{F}}}^{\rm{D}}
=\lim_{N\rightarrow\infty}{\mathcal{C}}_{\rm{M}}^{\rm{D}}=\{(0,0)\}.
\end{align}
This asymptotic result exhibits a different interpretation in the downlink case compared to the uplink counterpart interpreted in Remark \ref{remark_uplink_interpretation}, which is interpreted as follows. 
\vspace{-5pt}
\begin{remark}
As $N$ increases in the downlink PASS, the power allocated to each antenna, i.e., $\frac{P}{N}$, decreases. Therefore, the antennas that account for the majority of the power are positioned too far from the user, which makes the user receive negligible energy from the pinching antennas. 
\end{remark}
\vspace{-5pt}
Although the interpretations differ, the cause of this phenomenon is the same. The asymptotic degradation of the capacity region to zero is due to the fundamental nature of \emph{pinching beamforming}. This beamforming is a type of \emph{analog beamforming}, which lacks the capability to arbitrarily control the radiated power or enhance the effective signal power received from each antenna. Considering the results from both \eqref{Asymptotic_Rate_Capacity_Region_Uplink_Multiple} and \eqref{Asymptotic_Rate_Capacity_Region_Downlink_Multiple}, we derive the following key implication.
\vspace{-5pt}
\begin{remark}\label{Remark_Optimal_Antenna_Capacity_Region}
This finding implies that increasing the number of pinching antennas in either the uplink or downlink PASS does not lead to a monotonic expansion of the capacity/rate region. Instead, there exists an optimal number of antennas that maximizes overall system performance.
\end{remark}
\vspace{-5pt}
For convenience, Table \ref{Table: PASS_Downlink_Capacity_Region} summarizes the main results on the characterization of the downlink capacity and rate regions.
\section{Numerical Results}\label{Section_Numerical_Results}
We present numerical simulations to validate the analytical results. Unless stated otherwise, the system parameters are as follows: carrier frequency $f_{\rm{c}}=28$ GHz, effective refractive index $n_{\rm{eff}}=1.4$, one-dimensional search precision $Q=10^4$, minimum inter-antenna spacing $\Delta=\frac{\lambda}{2}$, and noise power $\sigma^2=-90$ dBm. The two users are uniformly distributed within a rectangular region centered at the origin, with side lengths $D_x=20$ m and $D_y=4$ m along the $x$- and $y$-axes, respectively, as illustrated in {\figurename} {\ref{Figure1}}. The waveguide is placed at a fixed height $d=3$ m along the $x$-axis with the $y$-coordinate given by $y_{\rm{p}}=0$ m. The $x$-coordinate of the feed point is set to $q_0=-\frac{D_x}{2}$ for the single-pinch case, and $q_0 =-\frac{D_x}{2}-1$ m for the multiple-pinch case. The maximum deployment range is set to $t_{\max}=\frac{D_x}{2}$ for the single-pinch case, and $t_{\max}=\frac{D_x}{2}+1$ m for the multiple-pinch case. As a benchmark, we compare against a conventional fixed-antenna system using a single antenna at location $[0,0,d]^{\mathsf{T}}$. All results are averaged over $10^3$ independent channel realizations.

\begin{figure}[!t]
\centering
    \subfigure[Capacity region.]
    {
        \includegraphics[width=0.4\textwidth]{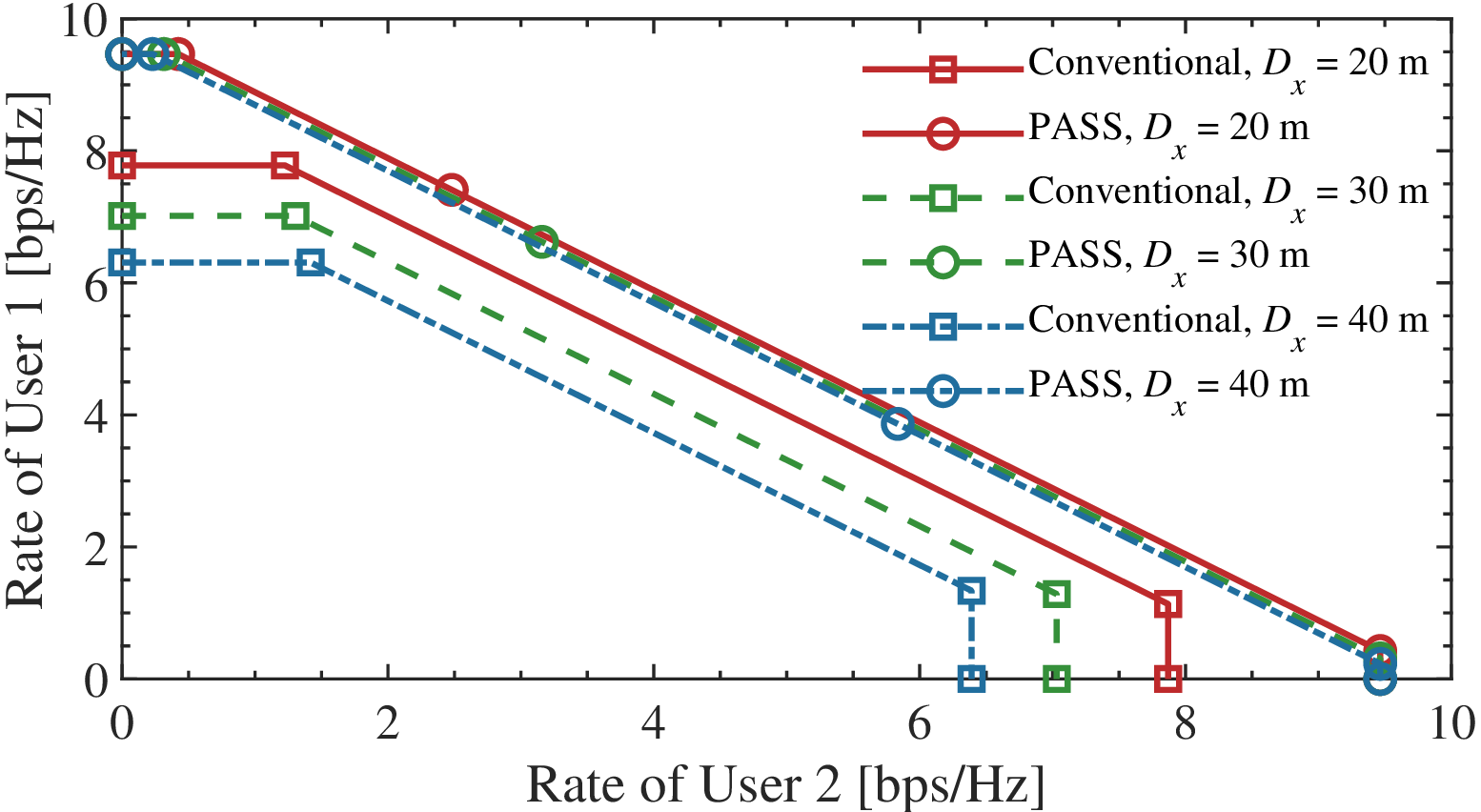}
	   \label{Figure_SP_MAC_Capacity_Range}
    }
    \subfigure[Achievable rate regions with TDMA and FDMA.]
    {
        \includegraphics[width=0.4\textwidth]{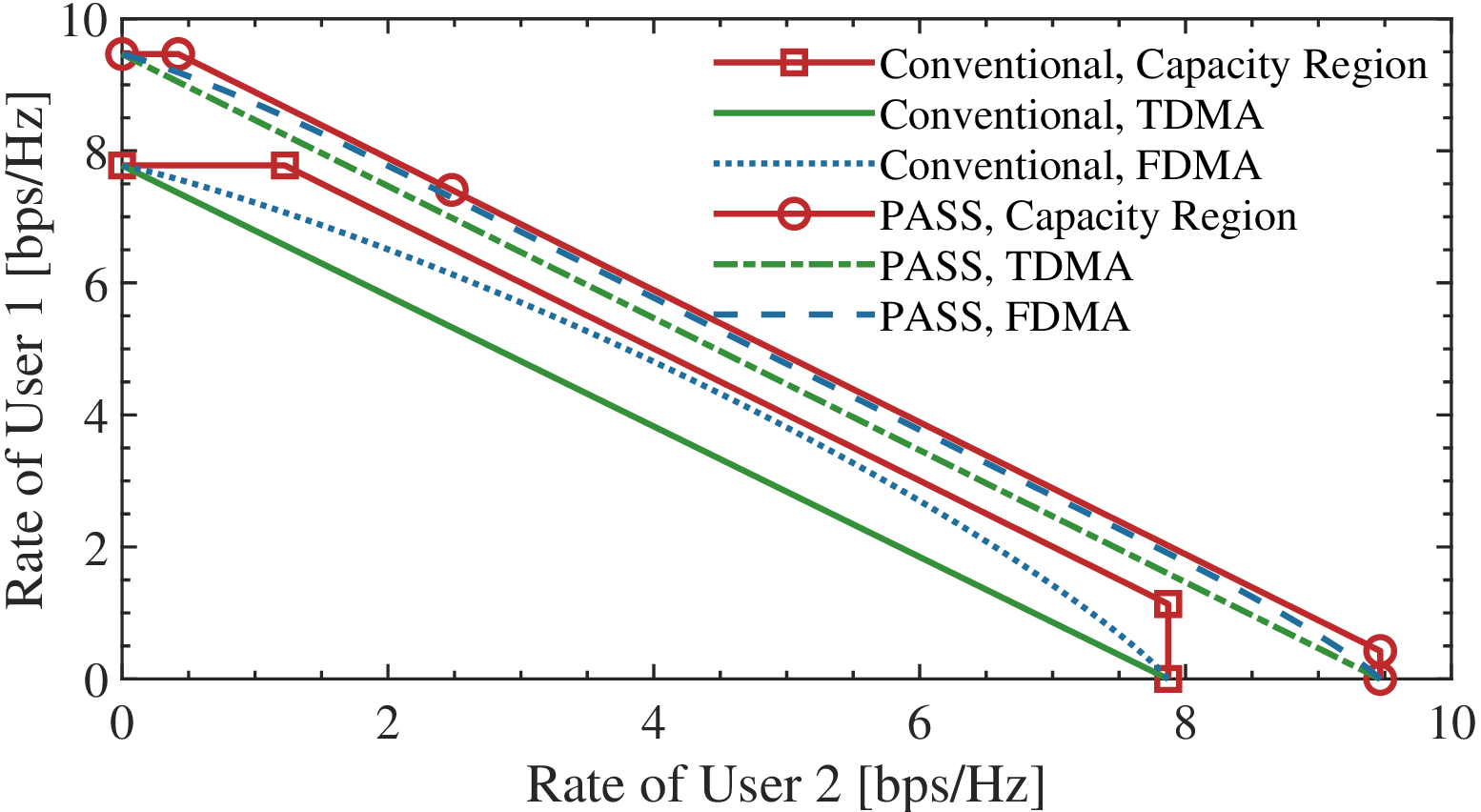}
	   \label{Figure_SP_MAC_Capacity_Rate_Region}
    }
\caption{Capacity/rate region comparison for PASS-enabled uplink channels in the single-pinch case.}
\label{Figure: MAC_Capacity_rate_region_comparison_Single_Pinch_PASS}
\vspace{-10pt}
\end{figure}

\subsection{PASS-Enabled Uplink Channels}
First, we focus on the uplink channel by setting the per-user transmit power as $P_1=P_2=10$ dBm, .
\subsubsection{Single-Pinch Case}
{\figurename} {\ref{Figure_SP_MAC_Capacity_Range}} compares the capacity regions of a conventional fixed-antenna system and a single-pinch PASS for different values of the side length $D_x$. As expected from Remark \ref{Capacity_Region_Comparision_Uplink}, the capacity region of PASS contains that of the fixed-antenna system. This thus validates the effectiveness of deploying PASS in enlarging the capacity region as well as the advantage of pinching beamforming design. Notably, the capacity gain becomes more pronounced as $D_x$ increases. This is because a larger $D_x$ increases the average distance between users and the center of the region, resulting in higher path loss for the fixed antenna. In contrast, the PASS can flexibly position its radiating element closer to the users, which effectively enhances the system throughput. As shown in {\figurename} {\ref{Figure_SP_MAC_Capacity_Range}}, while the conventional system's capacity region degrades substantially with increasing $D_x$, the degradation for PASS is negligible, which underscores the robustness and flexibility of pinching beamforming.

{\figurename} {\ref{Figure_SP_MAC_Capacity_Rate_Region}} illustrates the achievable rate regions for TDMA and FDMA under both the conventional fixed-antenna system and the single-pinch PASS. It is observed that for both systems, FDMA outperforms TDMA, which is consistent with our analytical results in Theorem \ref{Theorem_Comparision_TDMA_FDMA}. Furthermore, PASS yields a larger achievable rate region than the fixed-antenna system in both TDMA and FDMA scenarios. These results confirm that PASS outperforms conventional fixed-antenna system in both the capacity-achieving NOMA and practical OMA schemes.
\subsubsection{Multiple-Pinch Case}

\begin{figure}[!t]
\centering
    \subfigure[Capacity region.]
    {
        \includegraphics[width=0.4\textwidth]{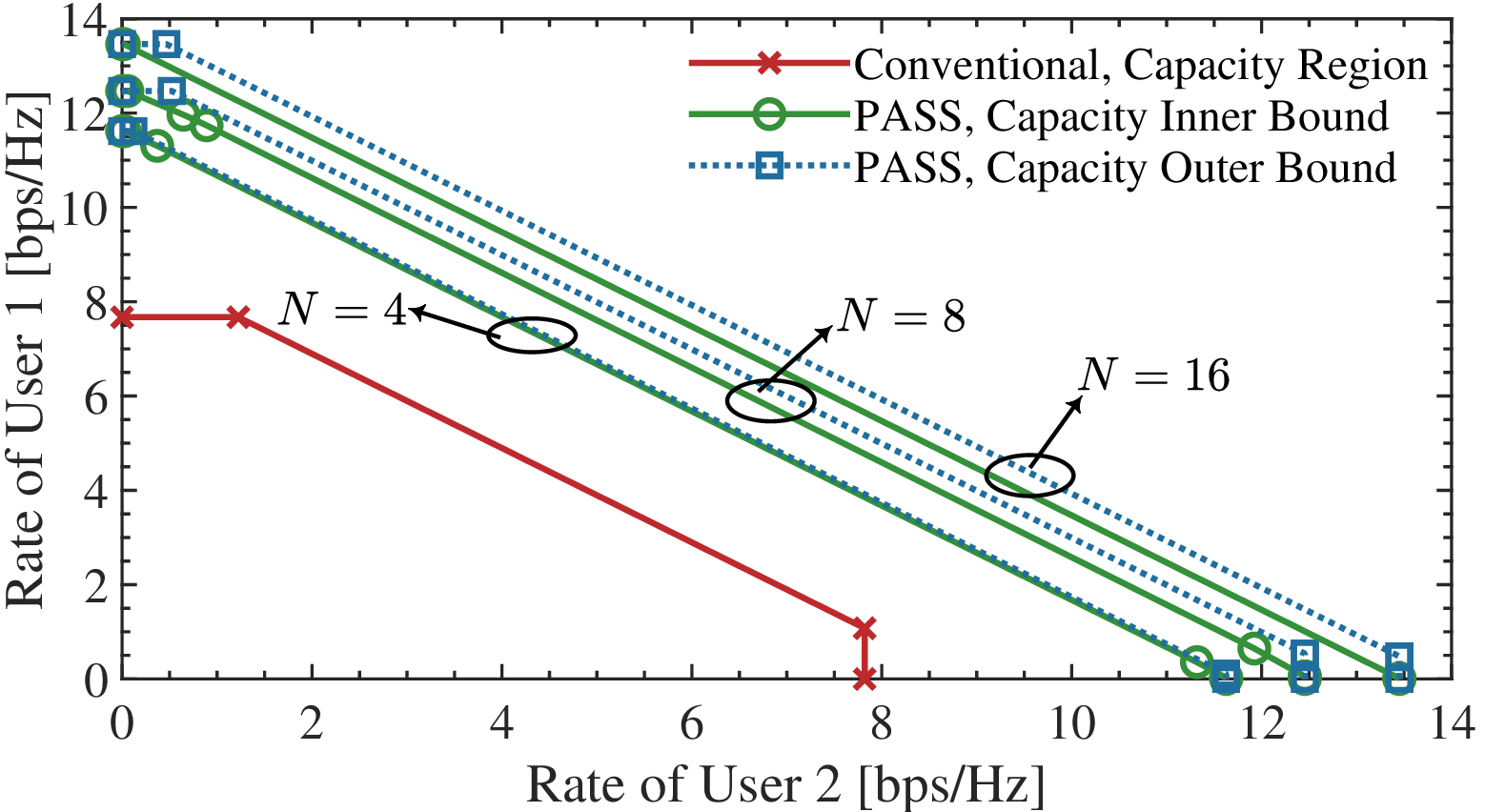}
	   \label{Figure_MP_MAC_Capacity_Rate_Antenna}
    }
    \subfigure[Achievable rate regions with TDMA and FDMA. $N=4$.]
    {
        \includegraphics[width=0.4\textwidth]{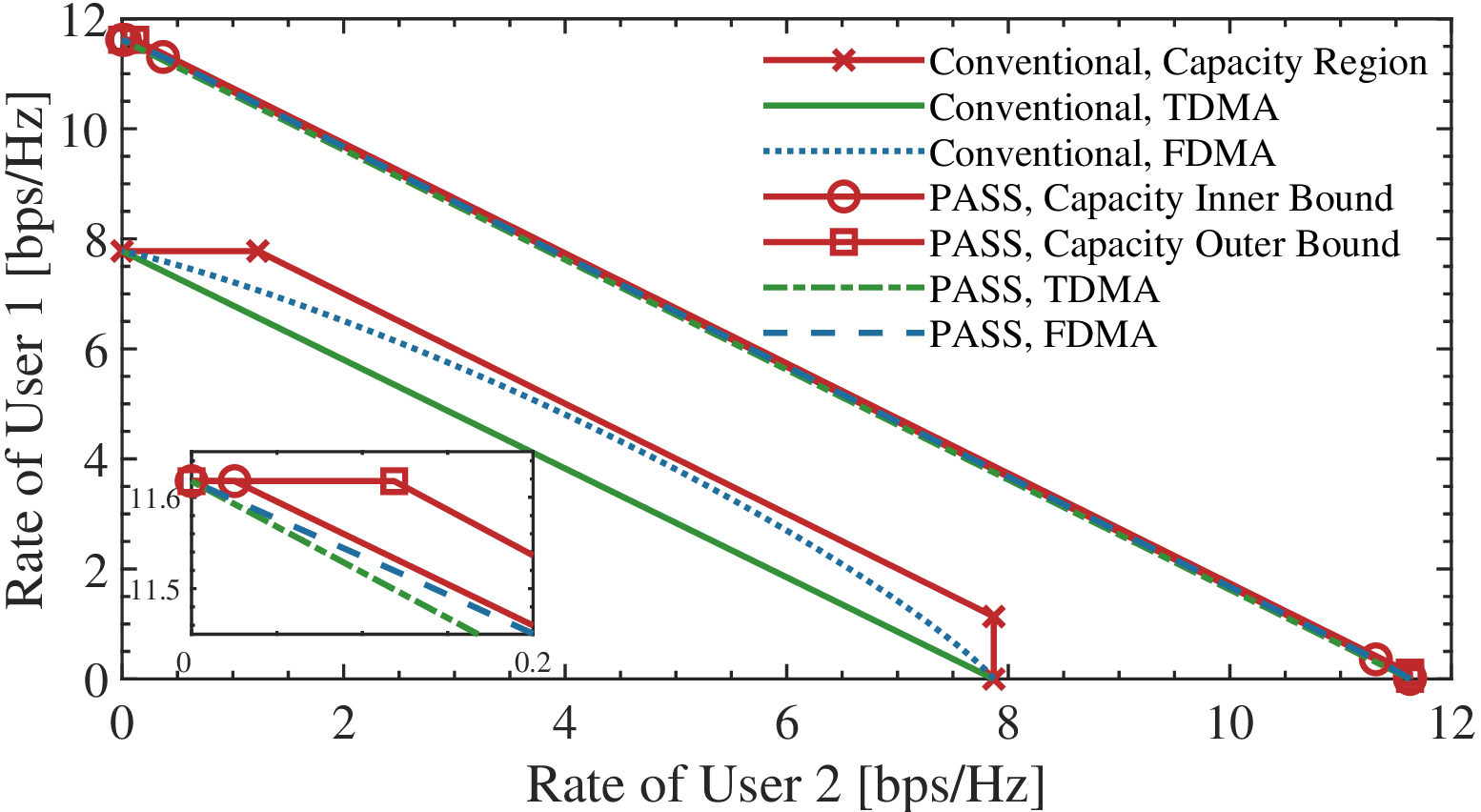}
	   \label{Figure_MP_MAC_Capacity_Rate_Region}
    }
\caption{Capacity/rate region comparison for PASS-enabled uplink channels in the multiple-pinch case.}
\label{Figure: MAC_Capacity_rate_region_comparison_Multiple_Pinch_PASS}
\vspace{-10pt}
\end{figure}

Next, we consider the multiple-pinch case. {\figurename} {\ref{Figure_MP_MAC_Capacity_Rate_Antenna}} compares the capacity regions of the conventional fixed-antenna system and the proposed multiple-pinch PASS. As shown, the derived inner and outer bounds closely align, which suggests that these bounds serve as good approximations for the exact capacity region. Since the inner bound is achievable using the proposed element-wise alternating optimization framework, the results in this graph validates its effectiveness in pinching beamforming design. The results also confirm that increasing the number of pinching antennas enlarges the capacity region. {\figurename} {\ref{Figure_MP_MAC_Capacity_Rate_Region}} presents the TDMA and FDMA achievable rate regions. It is observed that the comparison results among different regions are similar as those under the single-pinch case shown in {\figurename} {\ref{Figure: MAC_Capacity_rate_region_comparison_Single_Pinch_PASS}}, as expected. 

Notably, both subfigures reveal that the capacity region of multiple-pinch PASS is nearly \emph{triangular}. This shape arises because, at the two SIC corner points, the pinching beamformer is optimized for the second decoded user, which results in a high rate for that user and a seriously degraded rate for the first decoded user. Consequently, the overall region closely resembles the triangular TDMA region. This explains why the TDMA and FDMA achievable regions nearly coincide with the capacity region. We thus conclude that in multiple-pinch PASS, \emph{both TDMA and FDMA are nearly capacity-achieving}.

\begin{figure}[!t]
\centering
\includegraphics[width=0.4\textwidth]{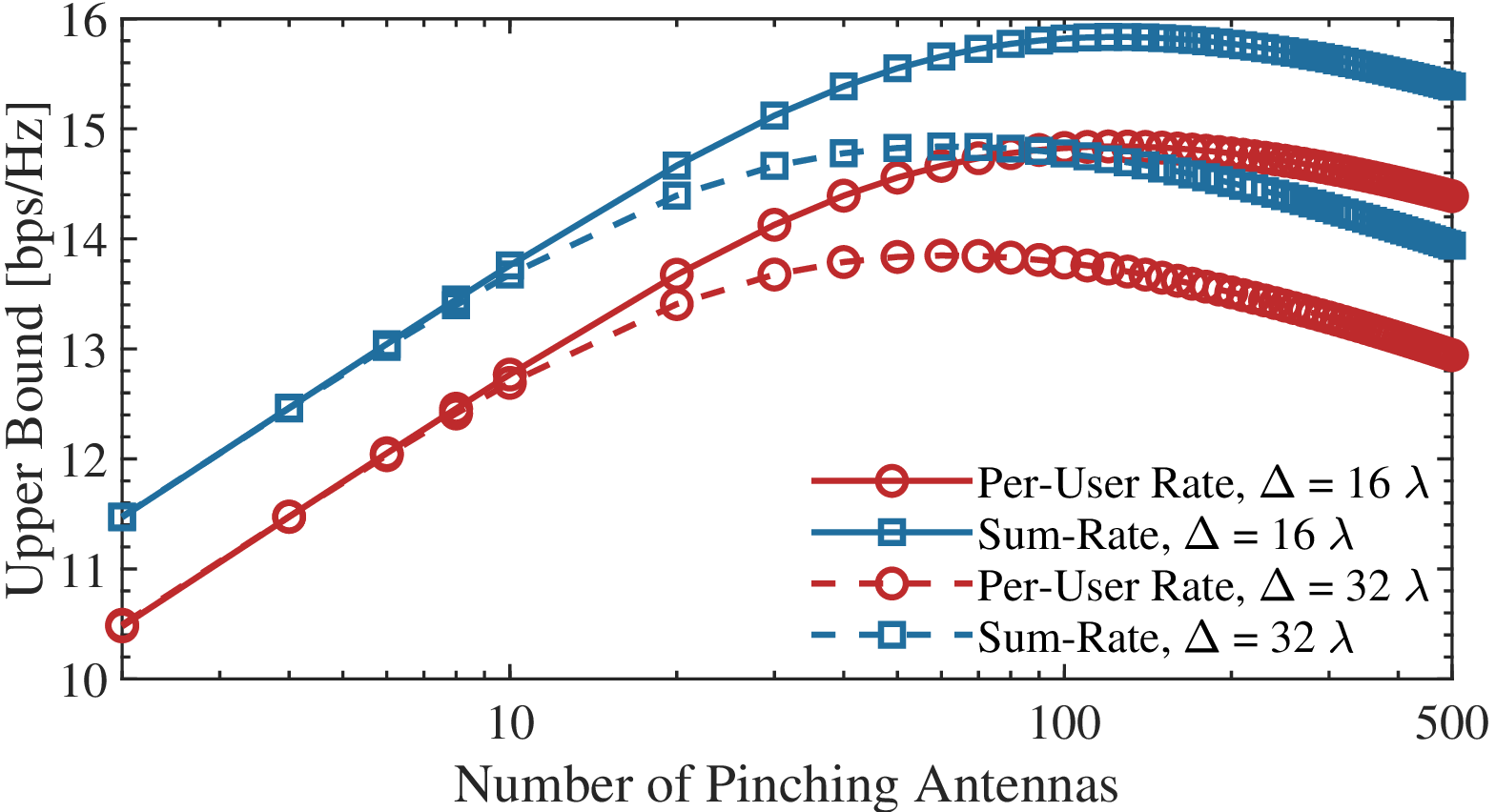}
\caption{Upper bounds of the per-user rate and sum-rate.}
\label{Figure_SP_Average_Rate_Range}
\vspace{-10pt}
\end{figure}

\begin{figure}[!t]
\centering
    \subfigure[Instantaneous capacity region. $x_1=4$ m, $x_2=6$ m, $y_1=1$ m, and $y_2=-2$ m.]
    {
        \includegraphics[width=0.4\textwidth]{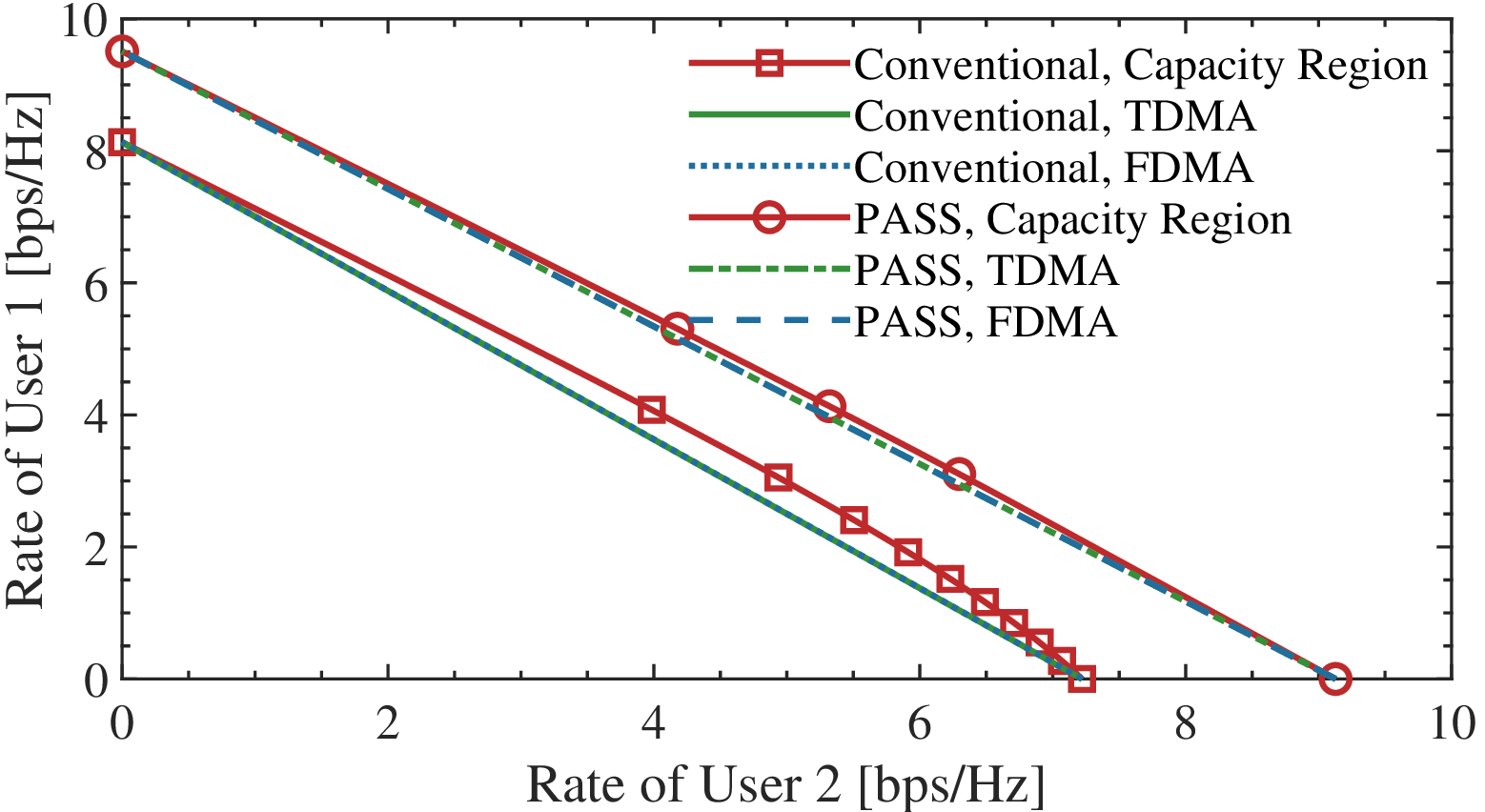}
	   \label{Figure_SP_Capacity_Range}
    }
    \subfigure[Average capacity region.]
    {
        \includegraphics[width=0.4\textwidth]{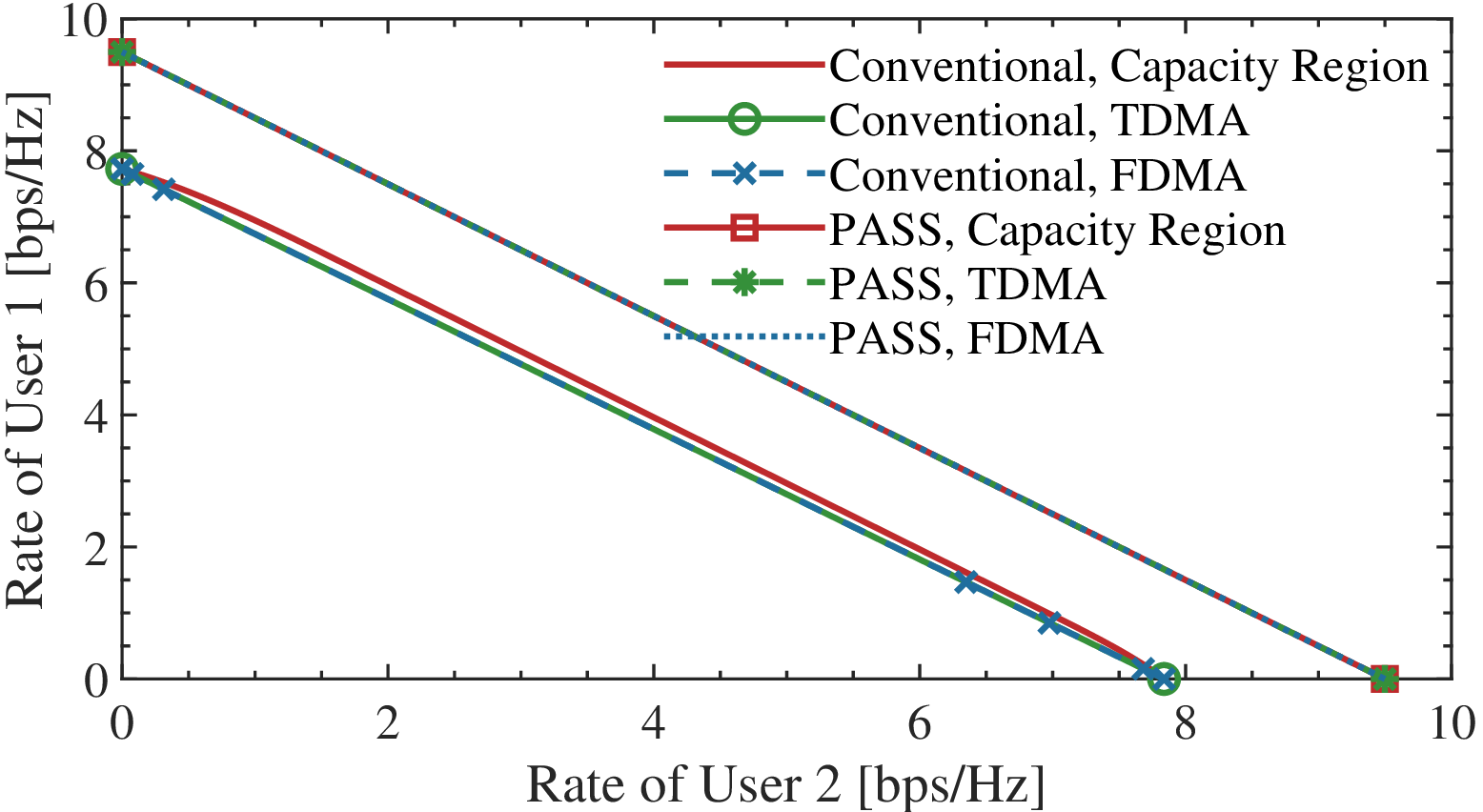}
	   \label{Figure_SP_BC_Capacity_Rate_Region}
    }
\caption{Capacity/rate region comparison for PASS-enabled downlink channels in the single-pinch case.}
\label{Figure: BC_Capacity_rate_region_comparison_Single_Pinch_PASS}
\vspace{-10pt}
\end{figure}

To investigate how the number of pinching antennas $N$ affects the capacity region, we fix the inter-antenna spacing $\Delta$ and plot the upper bounds of the per-user rate (\eqref{per_user_rate_bound}) and sum-rate (\eqref{sum_rate_upper_bound}) versus $N$ in {\figurename} {\ref{Figure_SP_Average_Rate_Range}}. The results show a non-monotonic trend: both bounds initially increase and then decrease with $N$, which validates our analytical results. As proven in \cite{ouyang2025array}, the bound of the per-user rate can be closely approximated by the refinement method in \cite{xu2024rate}. This thus suggests that there exists an optimal number of antennas that maximizes the capacity region, which validates the conclusion in Remark \ref{Remark_Optimal_Antenna_Capacity_Region}. For example, with $\Delta=16\lambda$, the optimal number of pinching antennas is approximately $120$, which spans an aperture of about $20$ m. This configuration is feasible for \emph{large-scale indoor deployments} (e.g., libraries, shopping malls). These findings underscore the importance of jointly optimizing both the \emph{number} of pinching antennas and their \emph{placement} strategy to approach capacity in practical large-scale PASS networks.

\begin{figure}[!t]
\centering
    \subfigure[Capacity region.]
    {
        \includegraphics[width=0.4\textwidth]{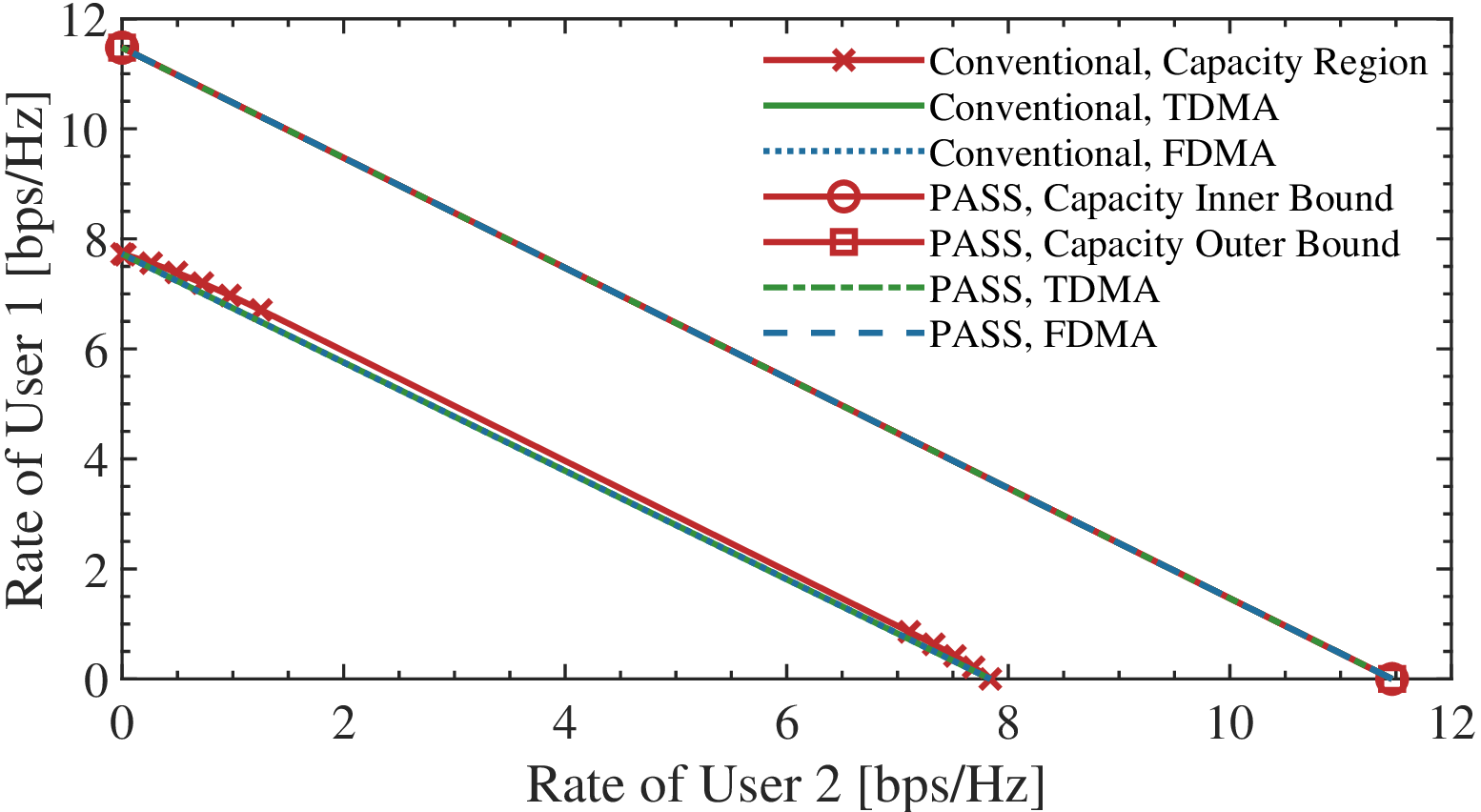}
	   \label{Figure_MP_BC_Capacity_Rate_Region}
    }
    \subfigure[Achievable rate regions with TDMA and FDMA.]
    {
        \includegraphics[width=0.4\textwidth]{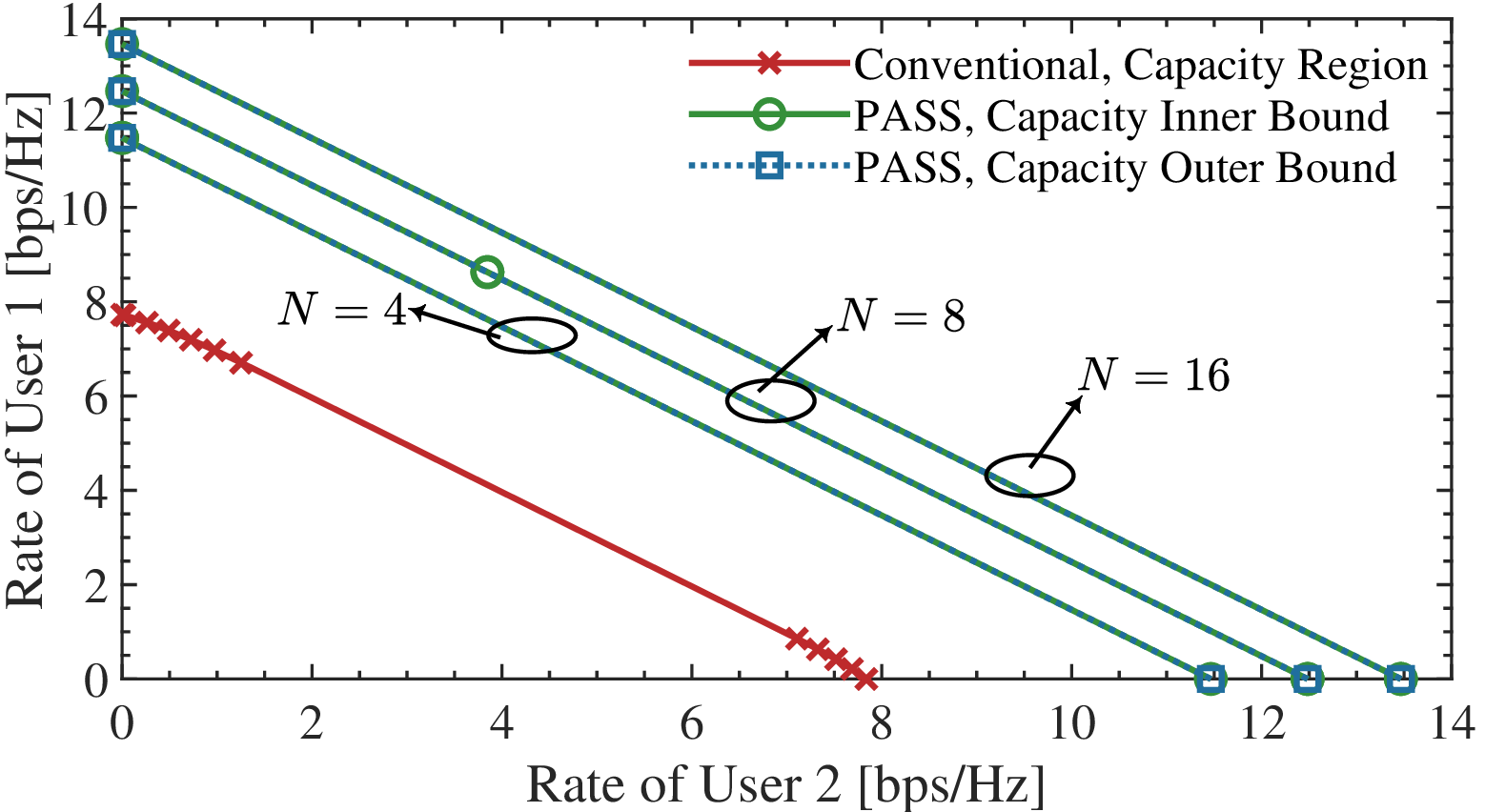}
	   \label{Figure_MP_BC_Capacity_Rate_Antenna}
    }
\caption{Capacity/rate region comparison for PASS-enabled downlink channels in the multiple-pinch case.}
\label{Figure: BC_Capacity_rate_region_comparison_Multiple_Pinch_PASS}
\vspace{-10pt}
\end{figure}

\subsection{PASS-Enabled Downlink Channels}
Next, we consider the downlink channel with $P=10$ dBm.
\subsubsection{Single-Pinch Case}
In {\figurename} {\ref{Figure_SP_Capacity_Range}} and {\figurename} {\ref{Figure_SP_BC_Capacity_Rate_Region}}, we compare the capacity and achievable rate regions of the fixed-antenna system and PASS for both an individual channel realization (\emph{instantaneous} case) and averaged over multiple realizations (\emph{average} case). In both scenarios, the capacity region of PASS encompasses that of the fixed-antenna system, validating our analysis in Section \ref{Section:PASS-Enabled Downlink Channels:Single-Pinch Case}. Specifically, in {\figurename} {\ref{Figure_SP_Capacity_Range}}, we observe that for PASS, the TDMA and FDMA achievable rate regions closely approximate the capacity region, while for the fixed-antenna system, they remain strictly suboptimal. This is due to the flexible positioning of the pinching antenna in PASS, which allows the channel gains of both users to be balanced even under the asymmetric deployment setting used for {\figurename} {\ref{Figure_SP_Capacity_Range}}. This flexibility results in a symmetric channel condition and a nearly \emph{triangular} capacity region, which TDMA and FDMA can effectively approximate through time sharing \cite[{\figurename} 6.6]{tse2005fundamentals}. This trend is further amplified in the average case shown in {\figurename} {\ref{Figure_SP_BC_Capacity_Rate_Region}}, where the average channel gains across users become more similar due to the adaptive antenna positioning in PASS. In contrast, the fixed-antenna system lacks this spatial adaptability, which limits the performance of TDMA and FDMA.

\subsubsection{Multiple-Pinch Case}
Finally, we consider the multiple-pinch downlink PASS. In {\figurename} {\ref{Figure_MP_BC_Capacity_Rate_Region}} and {\figurename} {\ref{Figure_MP_BC_Capacity_Rate_Antenna}}, we present the capacity and achievable rate regions for both the fixed-antenna system and the proposed PASS. It is observed that the derived capacity inner and outer bounds align with each other, validating their tightness. Another observation is that the achievable rate region attained by PASS fully encompasses the capacity region of the fixed-antenna system. As illustrated in both subfigures, increasing the number of pinching antennas leads to a significant expansion of the capacity region. Notably, for the multiple-pinch case, the achievable rate regions under TDMA and FDMA are found to closely approach the capacity region. This observation aligns well with our previous findings for the uplink case, which highlights the near-optimality of OMA schemes in practical multiple-pinch PASS deployments.
\section{Conclusion}\label{Section_Conclusion}
This article analyzed the information-theoretic capacity limits of PASS-enabled two-user uplink and downlink channels. We examined how pinching beamforming shaped the capacity region and the achievable rate regions under TDMA and FDMA. For the single-pinch case, we derived closed-form characterizations of these regions using the rate-profile approach. For the multiple-pinch case, we derived inner and outer bounds on the capacity region, as well as the TDMA and FDMA achievable data rate regions. We demonstrated the existence of an optimal number of pinching antennas that maximizes the capacity region. An important observation from our numerical results indicated that, in the multiple-pinch scenario, practical OMA schemes, including FDMA and TDMA, are capable of closely approximating the capacity region due to the spatial adaptability enabled by pinching beamforming. These findings highlight that, in practical PASS deployments, OMA schemes could be effectively utilized to achieve promising rate performance while reducing signal processing complexity in both encoding/decoding and pinching beamforming design.

The employed rate-profile-based framework for capacity region characterization could be extended to more general scenarios involving an arbitrary number of users by considering all possible decoding orders among users. Another important direction for further research is the study of the capacity region in the multiple-waveguide setting, where joint baseband and pinching beamforming designs would be required. 
\begin{appendix}
\subsection{Proof of Lemma \ref{Lemma_Single_PA_Basic_Step}}\label{Proof_Lemma_Single_PA_Basic_Step}
This result can be demonstrated via contradiction. If $q_1<x_1$, then $2x_1-q_1>q_1$. Updating $q_1\leftarrow 2x_1-q_1$ leaves $r_1^{\rm{U}}(q_1)$ unchanged but increases both $r_2^{\rm{U}}({{q}}_1)$ and $r_{1,2}^{\rm{U}}(q_1)$. If $q_1>x_2$, then $2x_2-q_1<q_1$. Setting $q_1\leftarrow 2x_2-q_1$ keeps $r_2^{\rm{U}}({{q}}_1)$ unchanged while improving $r_1^{\rm{U}}(q_1)$ and $r_{1,2}^{\rm{U}}(q_1)$. In both cases, a broader rate region ${\mathcal{C}}^{\rm{U}}(q_1)$ is achieved, which contradicts the optimality of $q_1\notin[x_1,x_2]$. Therefore, the capacity-achieving activated antenna location must lie within the interval $q_1\in[x_1,x_2]$, which yields ${\mathcal{C}}^{\rm{U}}={\rm{Conv}}(\bigcup\nolimits_{q_1\in[x_1,x_2]}{\mathcal{C}}^{\rm{U}}(q_1))$. This concludes the proof.
\subsection{Proof of Theorem \ref{Theorem_PASS_Uplink_Capacity_Region_SP_Closed-Form}}\label{Proof_Theorem_PASS_Uplink_Capacity_Region_SP_Closed-Form}
Given $\{\alpha,{\bm\pi}={\bm\pi}^{\rm{\uppercase\expandafter{\romannumeral2}}}\}$, the rate-profile problem formulated in \eqref{Rate_Profile_Uplink_Single_Pinch} seeks to maximize the following:
\begin{align}
q_{{\bm\pi}^{\rm{\uppercase\expandafter{\romannumeral2}}}}^{\alpha}=\argmax_{x\in[x_1,x_2]}\min\{f_{{\bm\pi}^{\rm{\uppercase\expandafter{\romannumeral2}}}}^{(2)}(x)/{\alpha},
f_{{\bm\pi}^{\rm{\uppercase\expandafter{\romannumeral2}}}}^{(1)}(x)/({1-\alpha})\}.
\end{align}
It is easy to verify that $f_{{\bm\pi}^{\rm{\uppercase\expandafter{\romannumeral2}}}}^{(1)}(x)$ (or $f_{{\bm\pi}^{\rm{\uppercase\expandafter{\romannumeral2}}}}^{(2)}(x)$) is strictly decreasing (or increasing) with $x\in[x_1,x_2]$. Then, if $\frac{1}{\alpha}f_{{\bm\pi}^{\rm{\uppercase\expandafter{\romannumeral2}}}}^{(2)}(x_1)>\frac{1}{1-\alpha}f_{{\bm\pi}^{\rm{\uppercase\expandafter{\romannumeral2}}}}^{(1)}(x_1)$, we have $q_{{\bm\pi}^{\rm{\uppercase\expandafter{\romannumeral2}}}}^{\alpha}=x_1$. If $\frac{1}{1-\alpha}f_{{\bm\pi}^{\rm{\uppercase\expandafter{\romannumeral2}}}}^{(1)}(x_2)>\frac{1}{\alpha}f_{{\bm\pi}^{\rm{\uppercase\expandafter{\romannumeral2}}}}^{(2)}(x_2) $, we have $q_{{\bm\pi}^{\rm{\uppercase\expandafter{\romannumeral2}}}}^{\alpha}=x_2$. Otherwise, $q_{{\bm\pi}^{\rm{\uppercase\expandafter{\romannumeral2}}}}^{\alpha}$ is determined by the point where $\frac{1}{\alpha}f_{{\bm\pi}^{\rm{\uppercase\expandafter{\romannumeral2}}}}^{(2)}(x)
=\frac{1}{1-\alpha}f_{{\bm\pi}^{\rm{\uppercase\expandafter{\romannumeral2}}}}^{(1)}(x)$ for $x\in[x_1,x_2]$, i.e., the intersection of the two functions. In this case, the optimal $q_{{\bm\pi}^{\rm{\uppercase\expandafter{\romannumeral2}}}}^{\alpha}$ can be obtained using a bisection search over the interval $x\in[x_1,x_2]$ to solve the equation $\frac{1}{\alpha}f_{{\bm\pi}^{\rm{\uppercase\expandafter{\romannumeral2}}}}^{(2)}(x)
=\frac{1}{1-\alpha}f_{{\bm\pi}^{\rm{\uppercase\expandafter{\romannumeral2}}}}^{(1)}(x)$. This concludes the proof.

\begin{figure}[!t]
\centering
\includegraphics[width=0.35\textwidth]{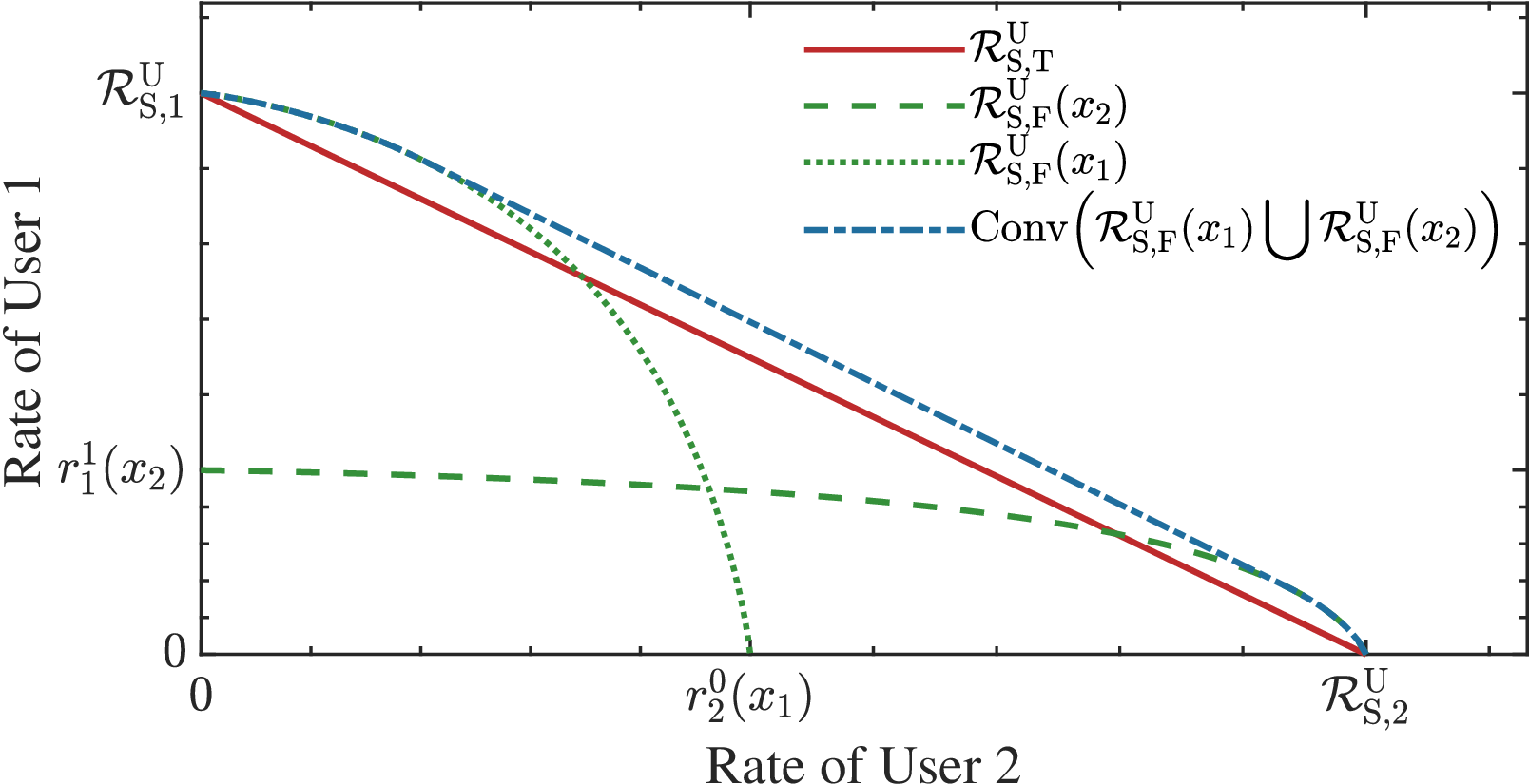}
\caption{Comparison of the TDMA and FDMA rate regions.}
\label{Figure_TDMA_FDMA_Region_AWGN}
\vspace{-10pt}
\end{figure}

\subsection{Proof of Theorem \ref{Theorem_Comparision_TDMA_FDMA}}\label{Proof_Theorem_Comparision_TDMA_FDMA}
Given $\{x_1,x_2\}$, the Pareto boundary of the TDMA achievable rate region ${\mathcal{R}}_{{\rm{S}},{\rm{T}}}^{\rm{U}}$ corresponds to a line segment connecting the points $(0,{\mathcal{R}}_{{\rm{S}},2}^{\rm{U}})$ and $({\mathcal{R}}_{{\rm{S}},1}^{\rm{U}},0)$, where $\log_2\left(1+\frac{P_k\eta}{d_k^2{\sigma^2}}\right)\triangleq{\mathcal{R}}_{{\rm{S}},k}^{\rm{U}}$ for $k\in\{1,2\}$, as shown in {\figurename} {\ref{Figure_TDMA_FDMA_Region_AWGN}}. Furthermore, the FDMA rate regions ${\mathcal{R}}_{{\rm{S}},{\rm{F}}}^{\rm{U}}(x_1)$ and ${\mathcal{R}}_{{\rm{S}},{\rm{F}}}^{\rm{U}}(x_2)$ are convex sets, which satisfies $r_{2}^{0}(x_1)\leq{\mathcal{R}}_{{\rm{S}},2}^{\rm{U}}$ and $r_{1}^{1}(x_2)\leq{\mathcal{R}}_{{\rm{S}},1}^{\rm{U}}$. Due to the curved Pareto boundaries of ${\mathcal{R}}_{{\rm{S}},{\rm{F}}}^{\rm{U}}(x_1)$ and ${\mathcal{R}}_{{\rm{S}},{\rm{F}}}^{\rm{U}}(x_2)$, it is straightforward to show that \emph{time sharing} between these two FDMA regions suffices to fully cover the TDMA region, as depicted in {\figurename} {\ref{Figure_TDMA_FDMA_Region_AWGN}}. Therefore, we conclude that
\begin{align}
{\mathcal{R}}_{{\rm{S}},{\rm{T}}}^{\rm{U}}\subseteq{\rm{Conv}}\Big({\mathcal{R}}_{{\rm{S}},{\rm{F}}}^{\rm{U}}(x_1)\bigcup{\mathcal{R}}_{{\rm{S}},{\rm{F}}}^{\rm{U}}(x_2)\Big)\subseteq{\mathcal{R}}_{{\rm{S}},{\rm{F}}}^{\rm{U}}.
\end{align}
This concludes the proof.
\subsection{Proof of Lemma \ref{Lemma_Array_Gain_Upper_Limitation}}\label{Proof_Lemma_Array_Gain_Upper_Limitation}
According to \cite{xu2024rate,ouyang2025array}, the maximization of the right-hand side of \eqref{Array_Gain_Upper_Bound_Maximization} is attained by uniformly placing the $N$ antennas with an inter-element spacing $\Delta$ and setting the center of the array aperture spanned by these antennas aligned with the projection of user $k$ along the waveguide, which yields
\begin{align}
q_n=(n-1)\Delta+q_1,\forall n>1,q_{N}+q_{1}=2x_k.
\end{align}
For notational simplicity, $N$ is assumed to be an even integer. In this case, we have 
\begin{subequations}
\begin{align}
A_{k}^{\triangleleft}&=\frac{\eta}{N}
\left\lvert\sum\nolimits_{n=1}^{{N}/{2}}\frac{2}{d_k\sqrt{1+(\Delta/2+(n-1)\Delta)^2/d_k^2}}\right\rvert^2\\
&=\frac{\eta}{N\Delta^2}
\left\lvert\sum\nolimits_{n=1}^{{N}/{2}}\frac{2\Delta/d_k}{\sqrt{1+((n-1/2)(\Delta/d_k))^2}}\right\rvert^2.\label{Array_Gain_Upper_Bound_Intermideate}
\end{align}
\end{subequations}
Since $\frac{\Delta}{d_k}\ll 1$, the summation in \eqref{Array_Gain_Upper_Bound_Intermideate} can be accurately approximated using a definite integral as follows:
\begin{subequations}
\begin{align}
A_{k}^{\triangleleft}&\approx\frac{\eta}{N\Delta^2}\bigg( \int_{0}^{{N\Delta}/{(2d_k)}}\frac{2}{\sqrt{1+x^2}}{\rm{d}}x\bigg)^2\\
&=\frac{4\eta\left(\ln\left(\sqrt{1+(\frac{N\Delta}{2d_k})^2}+\frac{N\Delta}{2d_k}\right)\right)^2}{N\Delta^2}.\label{Array_Gain_Upper_Bound_Later}
\end{align}
\end{subequations}
It follows from \eqref{Array_Gain_Upper_Bound_Later} that $A_{k}^{\triangleleft}\simeq{\mathcal{O}}\left({(\ln{N})^2}/{N}\right)$ as $N\rightarrow\infty$. This concludes the proof.
\end{appendix}
\bibliographystyle{IEEEtran}
\bibliography{mybib}
\end{document}